\documentclass[a4paper,11pt]{article}

\pdfoutput=1 

\usepackage{jinstpub} 

\usepackage{siunitx} 
\usepackage{hyperref} 
\usepackage{xcolor}
\usepackage{subcaption} 
\usepackage{multirow} 

\title{Design and Performance of the first IceAct Demonstrator at the South Pole}

\collaboration{The IceCube-Gen2 Collaboration}

\author[p]{M. G. Aartsen,}
\author[bh]{M. Ackermann,}
\author[p]{J. Adams,}
\author[l]{J. A. Aguilar,}
\author[t]{M. Ahlers,}
\author[ax]{M. Ahrens,}
\author[z]{C. Alispach,}
\author[am]{K. Andeen,}
\author[be]{T. Anderson,}
\author[l]{I. Ansseau,}
% ...

\emailAdd{analysis@icecube.wisc.edu}

\abstract{In this paper we describe the first results of IceAct, a compact imaging air-Cherenkov telescope operating in coincidence with the IceCube Neutrino Observatory (IceCube) at the geographic South Pole. An array of IceAct telescopes (referred to as the IceAct project) is under consideration as part of the IceCube-Gen2 extension to IceCube. Surface detectors in general will be a powerful tool in IceCube-Gen2 for distinguishing astrophysical neutrinos from the dominant backgrounds of cosmic-ray induced atmospheric muons and neutrinos: the IceTop array is already in place as part of IceCube, but has a high energy threshold. Although the duty cycle will be lower for the IceAct telescopes than the present IceTop tanks, the IceAct telescopes may prove to be more effective at lowering the detection threshold for air showers. Additionally, small imaging air-Cherenkov telescopes in combination with IceTop, the deep IceCube detector or other future detector systems might improve measurements of the composition of the cosmic ray energy spectrum. In this paper we present measurements of a first 7-pixel imaging air Cherenkov telescope demonstrator, proving the capability of this technology to measure air showers at the South Pole in coincidence with IceTop and the deep IceCube detector.}

\keywords{Cherenkov detectors, Gamma telescopes, Large detector systems for particle and astroparticle physics, Neutrino detectors}

\arxivnumber{1910.06945} 

\begin{document}

\maketitle
\flushbottom

\clearpage 

\newpage

\section{Introduction}
\label{sec:intro}

High-energy neutrinos are a unique probe to study the extreme high-energy universe. Neutrinos reach us from their production sites in the universe without absorption or deflection by magnetic fields.
The IceCube Neutrino Observatory has discovered a flux of high-energy neutrinos of cosmic origin
\cite{Aartsen:2014gkd,Aartsen:2016xlq}. The observed neutrino flux arrives almost isotropically at Earth. Recently, evidence for correlated neutrino and photon emission from the active galaxy TXS 0506+056 has been reported \cite{IceCube:2018dnn,IceCube:2018cha}.

IceCube's main instrument \cite{Aartsen:2016nxy} is a large-volume Cherenkov detector that instruments the glacial ice at the South Pole between depths from \SIrange{1.45}{2.45}{km} with \num{5160} digital optical modules
(DOMs) each containing a 10 inch photomultiplier tube \cite{Abbasi:2010vc} and associated electronics \cite{Abbasi:2008aa}. These DOMs are frozen into the ice along 86 vertical strings, with 60 DOMs per string.
In addition to the in-ice detector, the surface is instrumented with the IceTop air-shower detector \cite{IceCube:2012nn} that is composed of 162 Cherenkov tanks, each containing about 3 cubic meters of clear ice, instrumented with two DOMs.
The main purpose of this detector is the calibration of IceCube as well as cosmic ray physics in the energy range between the knee and ankle, roughly from \SIrange{e15}{e18}{eV}. Recently, also the capabilities to improve the sensitivity for astrophysical neutrino searches have been established by vetoing air-showers coincident with neutrinos and thus suppressing the background of cosmic ray induced muons \cite{Aartsen:2018vtx,Rysewyk:2019fdi}.

 As a result of the observations with IceCube,
the IceCube-Gen2 collaboration aims to substantially enhance the sensitivity of IceCube for astrophysical neutrino measurements \cite{Aartsen:2014njl,Aartsen:2015dkp}.
Three detector systems have been proposed to enhance the surface detector IceTop \cite{Tosi:2017zho} as potential extensions: a dense array of scintillator detectors IceScint \cite{Collaboration:2017tdy}, radio antennas \cite{V.:2017kbm} and an array of small air-Cherenkov telescopes (IceAct) that is subject of this paper. The purposes of IceAct, which overlap to varying degrees with the purpose of the other proposed extensions, include the following:

\begin{itemize}
	\item The coincident detection of cosmic ray-induced air showers and muons deep in the ice will allow for an improved calibration of the in-ice detector and IceTop.
	Direct measurements of the electromagnetic component of the air-shower with imaging air-Cherenkov telescopes can be compared to the high-energy muon component measured deep in the ice and the mixed component in IceTop. The independence of the telescopes from the ice and snow properties potentially provides a handle to reduce the influence of these systematic uncertainties \cite{Auffenberg:2017vwc,Ackermann:2017pja, Schaufel:2019aef}. The granularity of the camera of imaging air Cherenkov telescopes allows for precise measurements of cosmic ray-induced showers with very few telescopes.

	\item The observation of comic rays through several independent detection channels will also improve the capabilities of IceCube, IceTop, and IceAct to measure the composition of cosmic rays. High energy gamma-ray detection might also be possible \cite{Auffenberg:2017ypn}. A limited number of telescope stations to cover the overlap region of IceTop and IceCube in-ice promises a cost-efficient way to add an independent component to improve composition measurements in the energy range of IceCube in-ice and IceTop.

	\item The ability to veto high-energy muon events detected with IceCube, when a surface detector detects a coincident air-shower signal. This will reduce the atmospheric background events of muons and neutrinos in IceCube for diffuse neutrino measurements and point source searches in the southern sky \cite{Rysewyk:2019fdi, Auffenberg:2017vwc, Ackermann:2017pja, Schaufel:2019aef}.
	The low energy threshold of imaging air Cherenkov telescopes might greatly increase the sensitivity of IceCube in the southern sky for astrophysical neutrino detection down to \SI{30}{TeV} neutrino energy or lower with a dedicated array of IceAct telescopes \cite{Auffenberg:2017vwc, Schaufel:2019aef}. With a large enough sample of lower-energy astrophysical starting track events, it might also be possible to measure their inelasticity, and thus deduce the $\overline\nu:\nu$ ratio \cite{Aartsen:2018vez}.
	
\end{itemize}

The basic concept of the IceAct imaging air-Cherenkov telescopes is a compact and robust design, as outlined in \cite{Bretz:2018lhg}, optimized for operation in extreme environments and cost efficiency. Thus, the telescope has an enclosed optics with large field-of-view. This enclosure shields all delicate instrumentation from the harsh environment like the camera based on SiPMs.  Application of SiPMs in Cherenkov telescope was first demonstrated successfully in the First G-PAD Cherenkov (FACT) telescope \cite{2013JInst...8P6008A,2014JInst...9P0012B}.
In addition, the IceAct telescopes are much smaller than most imaging air-Cherenkov telescopes, with a diameter of \SI{55}{cm} and a tube length of about \SI{1}{m} including the DAQ, at the cost of a higher energy threshold of about \SI{10}{TeV} to \SI{20}{TeV} air-shower primary energy for a fully equipped 61 pixel telescope \cite{Schaufel2017,Rysewyk:2019fdi}. This is much higher than, for example, the small size telescopes planned for CTA \cite{SST} targeting a lower energy threshold. Due to the use of SiPMs for the camera and the enclosure of the entire optics (preventing us to get effected to much by scattered light) this instrument allows for a high duty cycle.\\ 
The energy threshold for air-shower detection of imaging air-Cherenkov telescopes is naturally much lower than achievable for particle detectors on the surface. The reason is that the air-Cherenkov light of air showers is predominantly emitted by the electromagnetic part of the air-shower during the entire air shower development. Compared to previous air Cherenkov non-imaging detection setups at the South Pole, like VULCAN \cite{VULCAN}, imaging air-Cherenkov telescopes can monitor the entire evolution of the particle cascade propagating through the atmosphere. In addition, the small field-of-view of the single pixels of the camera reduces the vulnerability to fluctuations in the night sky background light like auroras that distribute their light over large regions of the sky. 

\bigskip
This paper describes the design and performance of an initial IceAct demonstrator telescope with a 7-pixel camera. It was installed on the roof of the IceCube Laboratory at the South Pole in January 2016.
The goal of this demonstrator was to test the remote operation of a small imaging air-Cherenkov telescope under the harsh environment at the South Pole, and to verify the ability to observe air-showers in coincidence with IceCube and IceTop.
Based on the results that are described below, improved instruments with full field-of-view 61-pixel cameras, newly developed heating systems, and refined DAQ have been installed and commissioned in 2019. One of these is operating autonomously on the snow surface close to other air shower detector systems, while the other is deployed on the roof of the IceCube Laboratory.

\section{Hardware}

\begin{figure}[htbp]
	\centering
	\includegraphics[width=.69\textwidth]{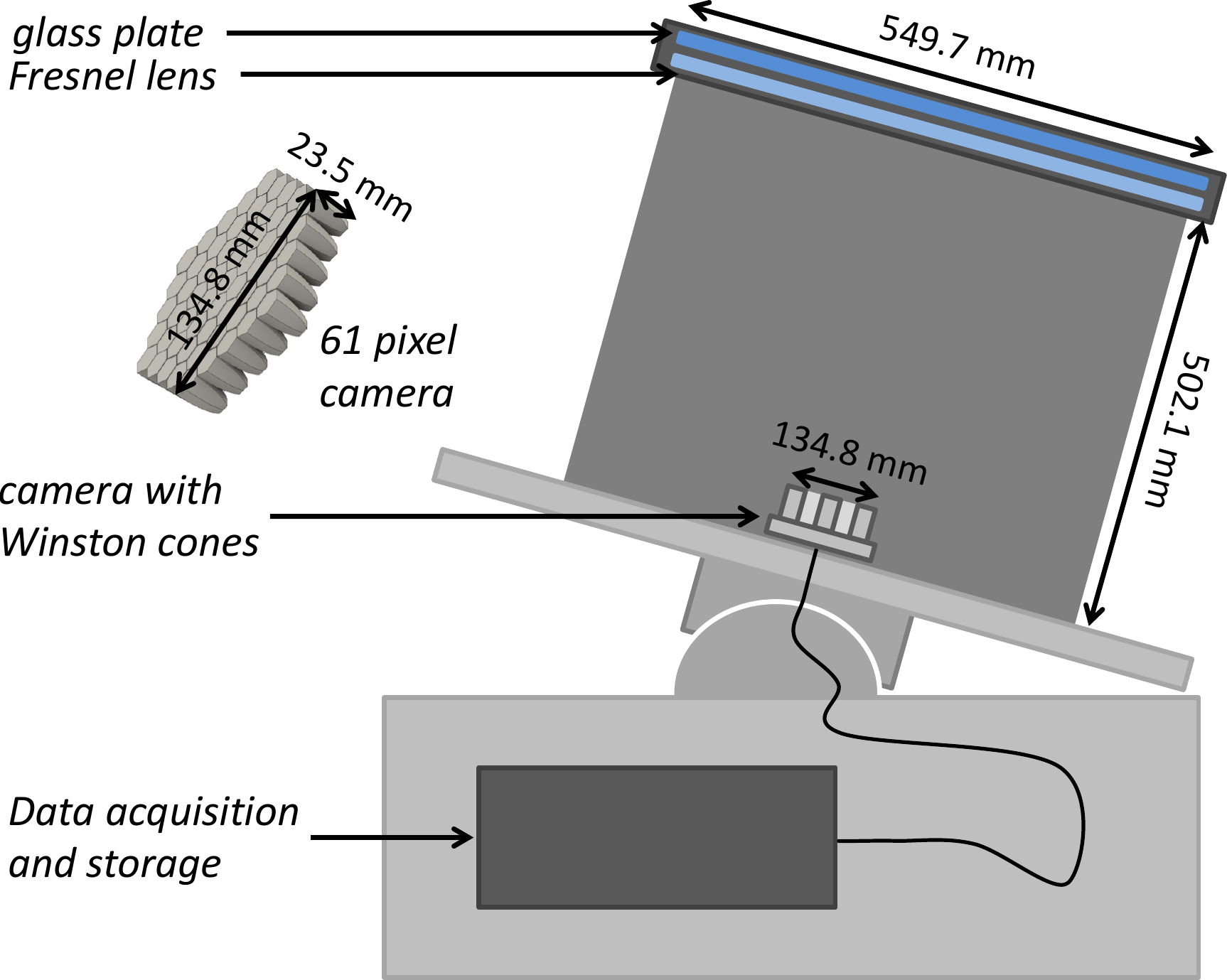}
	\caption{Drawing of the IceAct telescope design. The IceAct demonstrator of 2016 was equipped with a 7-pixel camera with $4^{\circ}$ field-of-view and was deployed at the South Pole on the roof of the ICL. In the figure a 61 pixel version of the camera is shown which is the current default design of IceAct.}
	\label{fig:telescope}
\end{figure}

The basic scheme of the demonstrator telescope is shown in Fig. \ref{fig:telescope}. The instrument follows the structure
in \cite{Bretz:2018lhg} with an enclosed tube of carbon-fiber and a Orafol SC943 Fresnel lens at the entrance window that is \SI{55}{cm} in diameter. The lens is protected from snow accumulation by a \SI{4}{mm} BOROFLOAT 33 glass plate.

\begin{figure}[htbp]
	\centering
	\includegraphics[width=.69\textwidth,origin=c,angle=0]{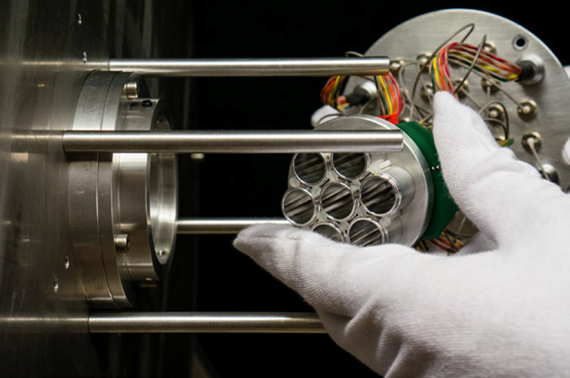}
	\caption{The 7-pixel camera of the IceAct demonstrator based on SensL-C SiPM sensors. The aluminum cones hold UG11 filters directly on the SiPMs.}
	\label{fig:camerahand}
\end{figure}

The camera in the focal point consists of seven SiPM pixels, each with \ang{1.5} field-of-view, as shown in Fig. \ref{fig:camerahand}.
The camera is equipped with SensL-FC SMT \SI{6}{mm} SiPMs that are mounted on a printed circuit board. Parabolic Winston cones made of aluminum with an embedded UG11 UV-light filter sit on top of each SiPM. The round light entrance window of the Winston cones have a diameter of $d_{in} = \SI{13.42}{mm}$ and their round exit window size fits to the SiPMs with $d_{out} = \SI{6.00}{mm}$.
The tube is fixed to a mount on top of a temperature-isolated wooden box which houses the slow-control and data acquisition electronics. A customized power supply \cite{Schumacher:2015dvx} provides a temperature-corrected bias-voltage to the SiPMs achieving a stable over-voltage. It offers a precision of \SI{1}{mV} and is connected via Ethernet to the slow-control computer. A switchable \SI{60}{W} heat source ensures that the temperature stays warm enough for stable operation. Two DRS4-evaluation boards connected via USB 2.0 to the slow control computer build the trigger and readout system of the data acquisition described in section \ref{sec:DAQ}.
The slow control software consists of two main parts: the web front-end and the back-end controlling the hardware components. The web interface allows users to control the entire telescope system and sorts the data in a MySQL database \cite{TimPHD}.

\begin{figure}[htbp]
	\centering
	\includegraphics[trim=90 30 30 0, clip, width=.69\textwidth]{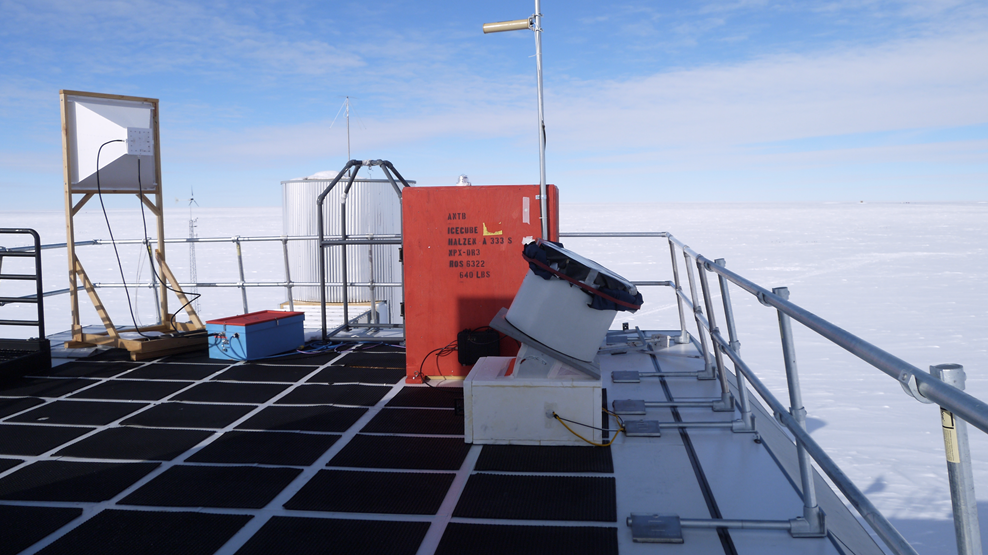}
	\caption{IceAct demonstrator on the rooftop of the IceCube Laboratory (ICL) at the geographic South Pole. The telescope is seen here tilted at an angle during installation, however during operation it was pointed straight upwards (within approximately \ang{1} of the vertical, see section \ref{sec:performance}).}
	\label{fig:roof}
\end{figure}

The instrument was installed at the South Pole in January 2016 on the roof of the IceCube Laboratory (ICL) as shown in Fig. \ref{fig:roof}, and operated during the austral winter pointing in the vertical direction. The overall telescope system weights about \SI{50}{kg}. The insulation box that houses all electronic components simultaneously acts as the stand for the telescope but was not fixed to the roof top of the ICL. The IceAct telescope was connected with \SI{110}{V} AC for power, RJ45 Ethernet cable for data transfer, and \SI{50}{\ohm} LEMO cable to distribute the trigger signal to IceCube. The entire DAQ system was located outside the building in the insulation box (see Figs. \ref{fig:telescope}, \ref{fig:roof}, and \ref{fig:logic}) to test all systems under the environmental stress (e.g. low temperature, low humidity, low pressure, continuous drift of micro snow crystals) at the South Pole. The temperature of the systems was continuously monitored to learn about possible failure modes. The complete telescope was operated remotely by an operator in the northern hemisphere via an ssh connection when internet connection was available during appropriate satellite passes. During that time, changes to the slow control and run settings could be made. The rest of the time, the telescope did not make any adjustments in response to changes in the environment, except for the temperature by correcting the SiPM bias-voltage.

\section{Data acquisition and triggering}
\label{sec:DAQ}

\begin{figure}[htbp]
	\centering
	\includegraphics[width=\textwidth]{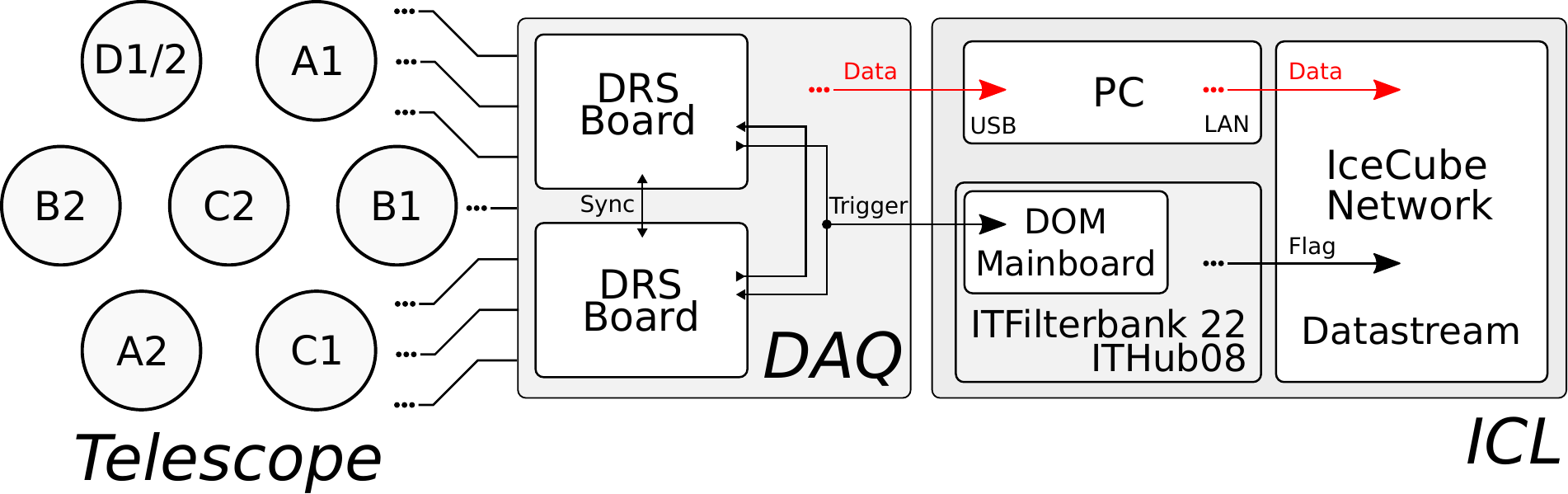}
	\caption{The DAQ of the IceAct demonstrator consists of two DRS4 evaluation boards \cite{DRSeval,DRS4} that trigger and digitize the camera signal. Trigger information is sent to an IceCube DOM main-board to produce trigger flags in the IceCube data stream. One of the pixels in the ring was read out by both DRS4 boards, the others by one channel each.}
	\label{fig:logic}
\end{figure}

The main parts of the telescope data acquisition hardware are two DRS4-evaluation boards
designed by the Paul-Scherrer-Institut \cite{DRSeval,DRS4}. These boards are responsible for the readout and digitalization of the SiPM signals. The sampling rate of the DRS-4 evaluation boards was set to \SI{1}{G Sa/s}, resulting in samples with $\Delta t = \SI{1}{\nano\second}$ and a total length of the recorded waveform of \mbox{$t_{\mathrm{waveform}} = 1024 \cdot \SI{1}{\nano\second} \approx \SI{1}{\micro\second}$} (due to 1024 storage cells for each input channel). Each board is connected to four pixels of the camera and records there signals with a dynamic range of \SIrange[retain-explicit-plus]{-500}{+500}{mV}. For calibration purposes, one pixel is connected with two boards (see Fig. \ref{fig:logic}). Both DRS4-evaluation boards can send their data through USB\,2.0 to a central mini-PC (NUC) whenever a trigger-decision is taken. The PC is connected to the IceCube network for further data access and for sending the data via satellite connection north (SPADE).

To connect the IceAct telescope to IceCube a DOM main-board \cite{Abbasi:2008aa} is used as shown in figure \ref{fig:logic}. Since these main-boards are used by all IceCube and IceTop DOMs it is easy to include an additional one into IceCube's DAQ system. DOM main-boards house all electronics needed for power supply and readout of a photo multiplier as well as electronics for time calibration with respect to the other DOMs \cite{Abbasi:2008aa}.
In case of the IceAct DOM main-board, there is no attached photo-multiplier but the IceAct external trigger output. Every time the telescope triggers (see below) the DOM main-board is triggered as well. This hit gets recorded by IceCube whenever there is a coincident, but independent, trigger of IceCube or IceTop. Therefore, IceAct, in its 2016 version, cannot trigger an IceCube detector readout.

The trigger decision of the demonstrator itself is taken between the two DRS4 boards. Each of the two DRS4-evaluation boards counts recorded waveforms with a signal below \SI{-40}{mV}. Three different conditions can cause a trigger:
\begin{itemize}
	\item In case one of the DRS4-evaluation board counts two or more channels within their time-over-threshold, a trigger-signal is sent to the DOM main-board and the other DRS4 evaluation board. Then both DRS4 boards request a readout by the mini-computer. 

	\item In case of only one channel above threshold in one DRS4-evaluation board, a trigger option (indicating one channel above threshold) is sent to the other DRS4 board. If the other board als has a signal over threshold within the next \SI{200}{\nano\second}, a trigger signal is sent to the DOM main-board and to the other DRS4 board. Again both boards request a readout by the mini-computer.

	\item Every \SI{17}{s} each DRS4 board enforces a fixed rate trigger (FRT) event.
\end{itemize}

\noindent
The first two are referred to as "physics triggers" and the third is the so-called "fixed rate triggers".

Each IceAct event consists of \num{1024} samples from each channel of both DRS4 evaluation-boards. It also comes with a timestamp in UTC format from the mini-PC and counter numbers from both DRS4 evaluation boards internal \SI{60}{MHz} clock, starting with 0 at the IceAct run start. In addition, the slow control records the temperature of each SiPM and the power supply, the average trigger-rate based on the last 10 events, and the trigger setting of the DRS4 boards. The waveforms are stored in ROOT format while the slow control data is stored in a MySQL database. The data is part of the IceCube Neutrino Observatory data stream \cite{Abbasi:2008aa}.

IceCube and IceTop detector events include a timestamp in UTC format from the IceCube timing system and have a typical length on the order of several \SI{10}{\micro\second}.
If an IceAct trigger arrived during an IceCube event (an event triggered by IceCube or IceTop), a flag is set in the IceCube data.

\section{South Pole Operation}

\begin{figure}[htbp]
	\centering
	\includegraphics[width=.75\textwidth]{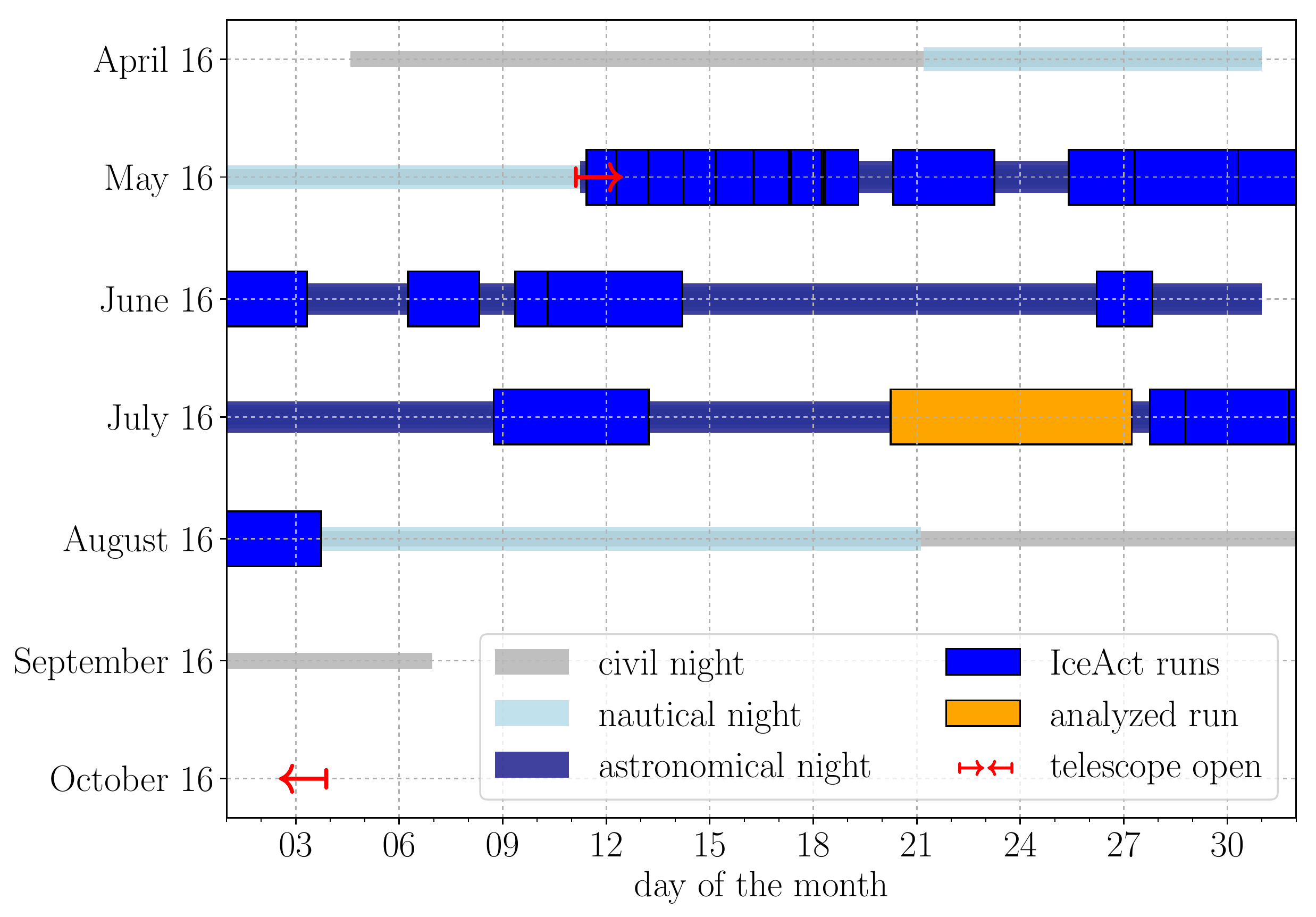}
	\caption{Up-time of the IceAct demonstrator in 2016 (blue and orange boxes). The astronomical night (sun \ang{18} below the horizon), during which most data was taken, was from May 12th to July 31st. The orange box indicates the data-taking period that is discussed in this paper. Civilian and nautical night correspond to the sun \ang{6} and \ang{12} below the horizon, respectively.}
	\label{fig:datataking}
\end{figure}

The IceAct demonstrator was operated pointing vertically during the austral winter of 2016 from May 12th to July 31st. Figure \ref{fig:datataking} shows the times when data was taken during this period. The orange block indicates the run that was chosen to be analyzed in this paper. This run, from July 20th to July 27th, is the longest consecutive run taken during the austral winter 2016 at South Pole. As the other runs have different settings to evaluate the system under South Pole conditions and due to technical tests, an analysis combining multiple runs was decided to be unnecessary for the scope of this paper. For example, the last run in Fig. \ref{fig:datataking} (July 31st to August 4th) is such a test run that was taken shortly after the end of the astronomical night.

\subsection{Environmental Conditions and cloud monitoring}

\begin{figure}[htbp]
	\centering
	\includegraphics[width=.75\textwidth]{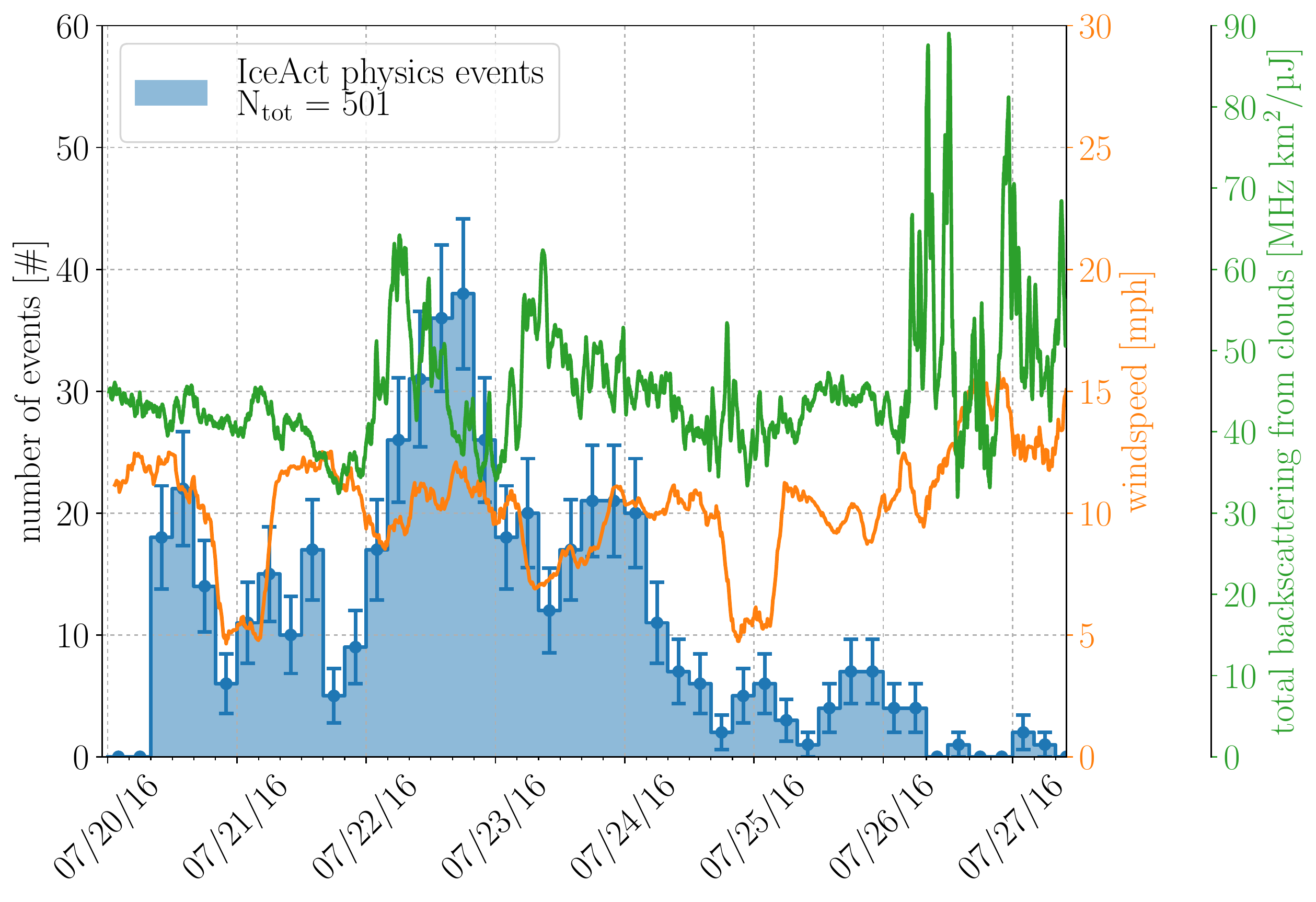}
	\caption{The number of physics events taken by IceAct is shown in blue. The orange curve show the backscattering of light from LIDAR measurements by MPLNET \cite{lewis2016overview,welton2001global}. The green line shows the wind-speed measured by the South Pole Weather station.}
	\label{fig:cloud}
\end{figure}

At the beginning of the long run on July 20th, 2016 (MJD 57589) there was, by chance, a full moon with \SI{99.8}{\percent} illumination decreasing from then on. On July 26th and 27th the moon was below the horizon. The moon was never more than \ang{16} above the horizon, which is safely outside of the field-of-view of the vertically-pointing telescope. No clear evidence of the light illumination by the moon was found. Figure \ref{fig:cloud} shows the physics trigger rate of the telescope in blue together with wind speed data in orange, and the total amount of back-scattered light from clouds measured by MPLNET \cite{lewis2016overview,welton2001global} at the South Pole in green. We see a possible correlation of the trigger rate with the amount of back-scattered light. The last days show a decrease in trigger rate of the telescope together with increasing back-scattering of light in the LIDAR data due to clouds, and also a slightly increased wind speed. A fish-eye camera was also installed close to the telescope and confirms the weather conditions found by the LIDAR.

However, this cannot explain all of the observed physics event rate variation. Additional snow accumulation on the entrance glass plate is likely to affect the observed rate. During four visits to the IceAct telescope during the austral night some slight amount of snow on the front glass plate was found to be easily removable with a brush. The snow accumulation in between these sparse visits is unknown. The exact amount of snow accumulation on the front glass is unknown during operation and will be further investigated with the next telescopes in the future. An automated wiper and a heating system are under development for the next generation telescopes. Here, power consumption and reliability will be critical parameters.

\subsection{Synchronization and Selection of Coincident Events with IceCube}

A comparison of IceCube and IceTop data with the data collected by the IceAct demonstrator demands a synchronization of the data streams. As described in section \ref{sec:DAQ}, a DOM main-board connected to IceCube was receiving trigger signals from the IceAct telescope. Whenever IceCube or IceTop were triggering, this DOM main-board was also read out. If the main-board has received an IceAct trigger, then this would result in an IceAct trigger flag in the IceCube data stream. This allows for a synchronization of the timestamp in the IceAct data (from the mini PC in UTC format) with the GPS-synchronized UTC time of the IceCube detector system. Even though before the synchronization it is unknown which IceAct event is coincident with IceCube, we know that each trigger flag in IceCube corresponds to one particular IceAct event. Therefore, the time pattern of the IceAct events and the time pattern of the IceCube events with an IceAct trigger flag can be matched up by finding a constant time offset. This brings the time-distribution patterns of the two sets of events into alignment. However, if there are multiple telescope runs to be analyzed it needs to be applied for each run individually.

\begin{figure}[htbp]
    \centering
    \includegraphics[width=.75\textwidth]{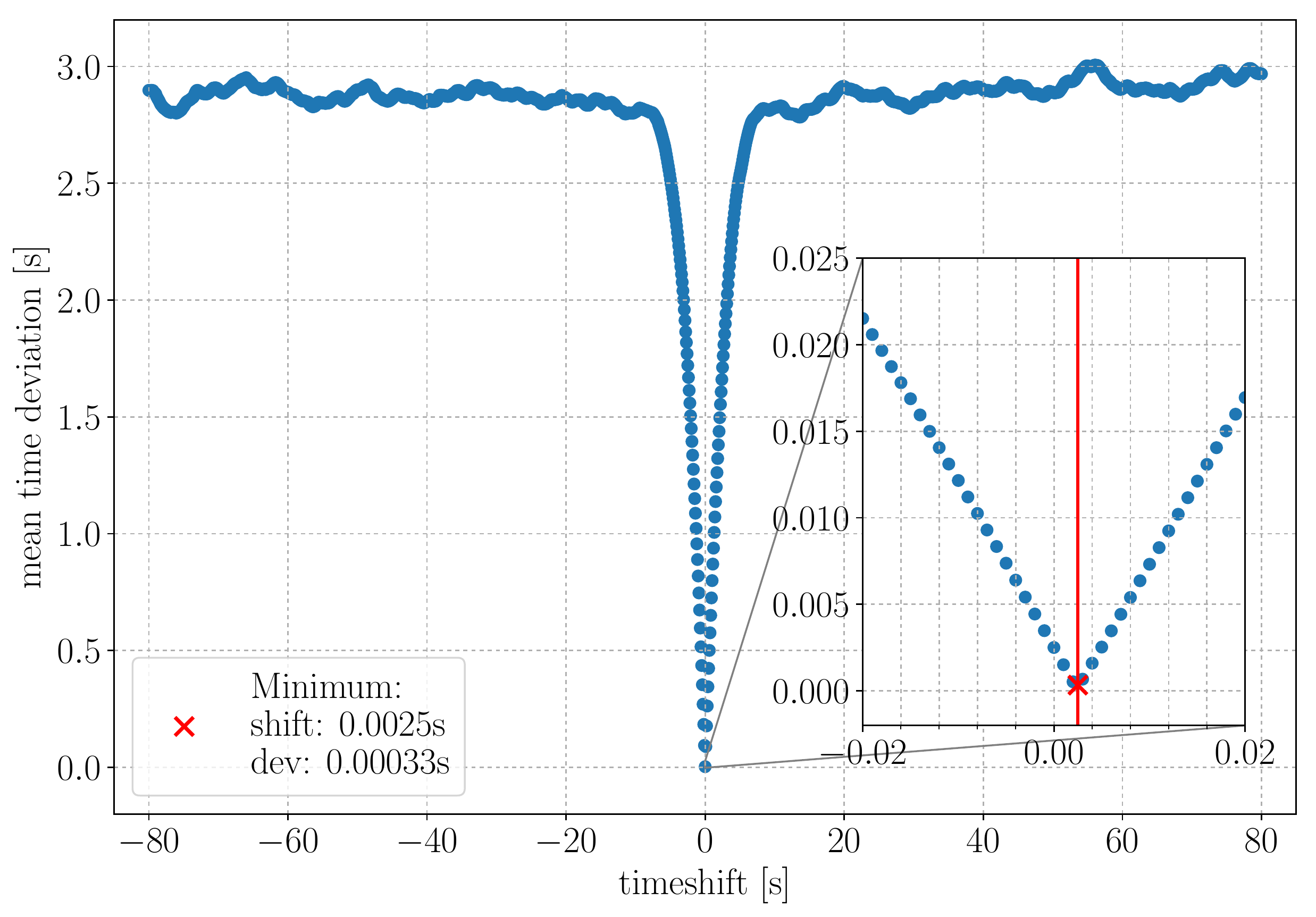}
    \caption{Alignment of IceCube and IceAct events using physics triggers of the telescope only. Shown is the mean time deviation between next-neighbor IceAct and IceCube events as a function of the time-shift between both event streams. The absolute minimum indicates the optimum time-shift between the IceCube and IceAct timestamps. The mean time deviation found for this optimum time-shift between both event streams is called 'dev'.}
	\label{fig:alignment}
\end{figure}

This alignment process is performed as follows: first, the mean time deviation between the closest event pairs is calculated for a fixed offset between both sets of events. In a second step this offset is varied in the range of \SIrange[retain-explicit-plus]{-100}{+100}{s} in steps of \SI{1}{\milli\second} assuming that the IceAct and the IceCube event times do not have a difference larger than \SI{100}{s}. The shift yielding the least mean time deviation then corresponds to the offset between both sets. In Fig. \ref{fig:alignment}, the mean time deviation of all IceCube events relative to the IceAct events closest in time is shown as a function of the time offset between both data-streams.

This time synchronization method was applied twice: first with all IceAct and IceCube events (that have an IceAct trigger flag) and secondly with events triggered by the IceAct FRT filtered out. The filtering of FRT triggered events requires a successful time synchronization in the first step, since the trigger flag in IceCube does not differ for FRT and physics triggered events, but can be used as a cross-check. In both synchronizations, including fixed rate triggers or not, a clear minimum of the mean time deviation can be found for a best fit time-shift of $\Delta t = \SI{2.5}{\milli\second}$. This corresponds to the optimum time-shift between IceAct and IceCube for this particular run. Figure \ref{fig:alignment} shows the distinct minimum that is found using only physics triggered events.

\begin{figure}[htbp]
	\centering
	\includegraphics[width=.75\textwidth]{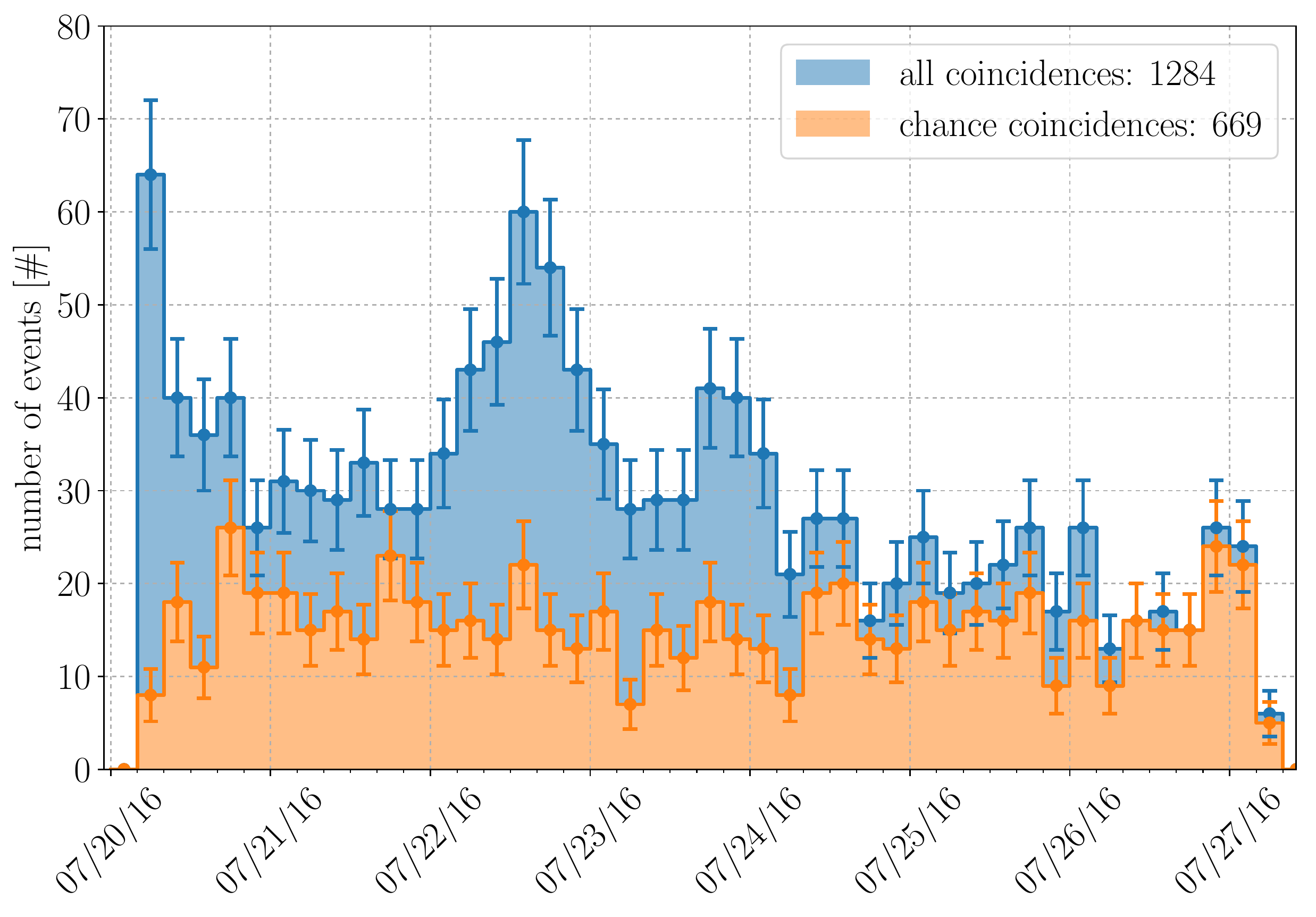}
	\caption{Histogram of events triggered by IceAct from July 20\textsuperscript{th} to July 27\textsuperscript{th}, 2016. The orange events are fixed rate triggers while the blue events are real (physics) triggers. The overall distribution is influenced by environmental conditions like the phases of the moon, clouds, and snow.}
	\label{fig:coincrates}
\end{figure}

The IceAct run that was analyzed contains \num{77810} events in total. We can select \num{1284} coincident IceAct, IceTop, and IceCube events between July 20th and July 27th, \num{615} of them caused by a physics trigger. The rate of identified coincident events is shown in Fig. \ref{fig:coincrates} and further summarized in table \ref{tab:IceActEvents}. After re-triggering by software and requiring a \SI{-40}{mV} threshold to remove events in the beginning of the run where the trigger threshold of the system was not yet set properly, \num{567} events remain. A \SI{333}{Hz} cut is applied to account for the maximum readout and trigger rate of the DAQ system. As one pixel of the camera is read out by both DRS4 boards there are events that are only seen by one pixel but in two channels. These events are filtered for further analyses by requiring more than two camera pixels causing a signal over threshold (Pixel multiplicity $\geq$ 2). In the case of a fixed rate trigger this filter is skipped. We require the time synchronization between the IceCube events and IceAct events to have been successful on an event to event basis (time deviation < \SI{3}{ms}) and, lastly, the RMS of the baseline of the channels to be smaller than \SI{8}{mV} to cut on events with electromagnetic interference. An overview of event numbers after only applying each single filter to the coincident events is given in table \ref{tab:IceActEvents}.

The filtered data now includes two uncorrelated types of coincidences: random coincident events of fixed rate triggers of IceAct, and physics coincidences. As the fixed rate trigger coincidences are chance coincidences only, they show an approximately constant rate over the observation time (see Fig. \ref{fig:coincrates}). This reflects the stable operation of both systems, IceCube and IceAct, during the whole run. The physics coincidences, however, show a strongly varying rate, reflecting changing cloudiness of the sky and potential snow accumulation on the telescope towards the end of the run (see also Fig. \ref{fig:cloud}).

Table \ref{tab:IceTopIceCubeEvents} summarizes the number of events with a coincident time stamp of IceTop and/or IceCube events together with IceAct. The events were split in two data-sets: one containing IceAct-IceTop and the other one containing IceAct-IceCube coincidences. These data-sets are not independent and there is an overlap of \num{123} events between them.

For the events in coincidence with IceTop, standard quality cuts are chosen to ensure a stable reconstruction by IceTop \cite{IceCube:2012nn}. First, the following cuts are applied: at least five IceTop stations need to contribute to the trigger decision; the IceTop event reconstruction of the air shower direction needs to have been successfully converged; and the event needs to be reconstructed to be contained in the IceTop footprint. In addition, the largest signal must be $\geq$ \SI{6}{VEM} (VEM: the signal that a single muon that vertically passes an IceTop tank produces in IceTop) and one of the stations next to the station with the largest signal needs a signal $\geq$ \SI{4}{VEM} ($S_{\mathrm{max}} > \SI{6}{VEM}$ and $S_{\mathrm{max}}^{\mathrm{neighbor}} > \SI{4}{VEM}$ in table \ref{tab:IceTopIceCubeEvents}). 320 events in coincidence between IceTop and IceAct remain for further analysis.

Standard cuts are applied for events in coincidence between IceAct and the IceCube in-ice array to ensure the stability of in-ice based directional muon reconstructions and a time correlation of the events \cite{IceCube:EnergyReco}. The hit time of an IceCube in-ice event has to be \SIrange{5}{8}{\micro\second} after the IceAct event time, as the particles are reaching the deep ice detector with the speed of light. In addition the directional reconstruction of the IceCube event needs to be successful (reduced Likelihood of MPEfit < 10) (fit quality in table  \ref{tab:IceTopIceCubeEvents}). Also the energy reconstruction has to be successful (energy in table  \ref{tab:IceTopIceCubeEvents}). The number of DOMs with a direct hits needs to be larger than \num{35} where a direct hit of a DOM is defined as a hit in agreement with the expected arrival time of the Cherenkov light of the reconstructed event within \SIrange[retain-explicit-plus]{-15}{+125}{\nano\second}. These direct hits must spread across at least two strings. After all cuts, 130 in-ice events in coincidence with the IceAct telescope remain.

\begin{table}[htbp]
	\caption{Summary of applied IceAct quality cuts. 501 events survive all quality cuts.}
	\label{tab:IceActEvents}
	\centering
	\begin{tabular}{c|c|S|S}
		\multirow{2}{*}{cut} & \multirow{2}{*}{setting} & \multicolumn{2}{c}{events passing} \\
	     & & {single cut} & {cumulative} \\ \hline \hline
		\multicolumn{2}{c|}{total number of coincident events} & 1284 & 1284 \\ \hline
		\multicolumn{2}{c|}{number that are physics-triggered} & 615 & 615 \\ \hline
		Trigger threshold & $< \SI{-30}{\milli \volt}$ & 567 & 567 \\ \hline
		Trigger rate & $< \SI{333}{\hertz}$ & 614 & 567 \\ \hline
		Pixel multiplicity & $\geq \num{2}$ & 521 & 521 \\ \hline
		Time synchronization & $< \SI{3}{\milli \second}$ & 601 & 508 \\ \hline
		Baseline RMS & $< \SI{8}{\milli \volt}$ & 606 & 501
	\end{tabular}
\end{table}

\begin{table}[htbp]
	\centering
	\caption{Summary of applied IceTop and IceCube in-ice quality cuts for events in coincidence with IceAct triggers.}
	\label{tab:IceTopIceCubeEvents}
	\begin{subtable}[t]{0.6\textwidth}
		\subcaption{IceTop quality cuts}
		\begin{tabular}{c|c|S|S}
			\multirow{2}{*}{cut} & \multirow{2}{*}{setting} & \multicolumn{2}{c}{events passing} \\
			 & & {single cut} & {cumulative} \\ \hline \hline
			\multicolumn{2}{c|}{total number of events} & 426 & 426 \\ \hline
			$N_{\mathrm{Station}}$ & $\geq 5$ & 424 & 424 \\ \hline
			fit quality & true\footnote{air shower reconstruction must have converged} & 393 & 393 \\ \hline
			containment & true\footnote{impact point must lie within the IceTop footprint} & 423 & 392 \\ \hline
			$S_{\mathrm{max}}$ &  $> \SI{6}{VEM}$ & 408 & 383 \\ \hline
			$S_{\mathrm{max}}^{\mathrm{neighbor}}$ & $> \SI{4}{VEM}$ & 355 & 339 \\ \hline \hline
			\multicolumn{2}{c|}{with IceAct cuts} & \multicolumn{2}{S}{320}
		\end{tabular}
	\end{subtable}
	\vfill
	\begin{subtable}[t]{0.6\textwidth}
		\subcaption{In-ice quality cuts}
		\begin{tabular}{c|c|S|S}
			\multirow{2}{*}{cut} & \multirow{2}{*}{setting} & \multicolumn{2}{c}{events passing} \\
			 & & {single cut} & {cumulative} \\ \hline \hline
			\multicolumn{2}{c|}{total number of events} & 240 & 240 \\ \hline
			timing & $\SI{5}{\micro\second} < \Delta T < \SI{8}{\micro\second}$ & 194 & 194 \\ \hline
			fit quality & $\mathcal{L}_{\mathrm{reduced}} < 10$ & 223 & 194 \\ \hline
			energy & true\footnote{reconstruction must be successful} & 220 & 188 \\ \hline
			$N_{\mathrm{DOMs}}^{\mathrm{direct}}$ & $> 35$ & 152 & 150 \\ \hline
			$N_{\mathrm{strings}}^{\mathrm{direct}}$ & $> 1$ & 213 & 134 \\ \hline \hline
			\multicolumn{2}{c|}{with IceAct cuts} & \multicolumn{2}{S}{130}
		\end{tabular}
	\end{subtable}
\end{table}

\begin{figure}[htbp]
	\centering
	\begin{subfigure}[t]{0.49\textwidth}
		\centering
		\includegraphics[width=\textwidth]{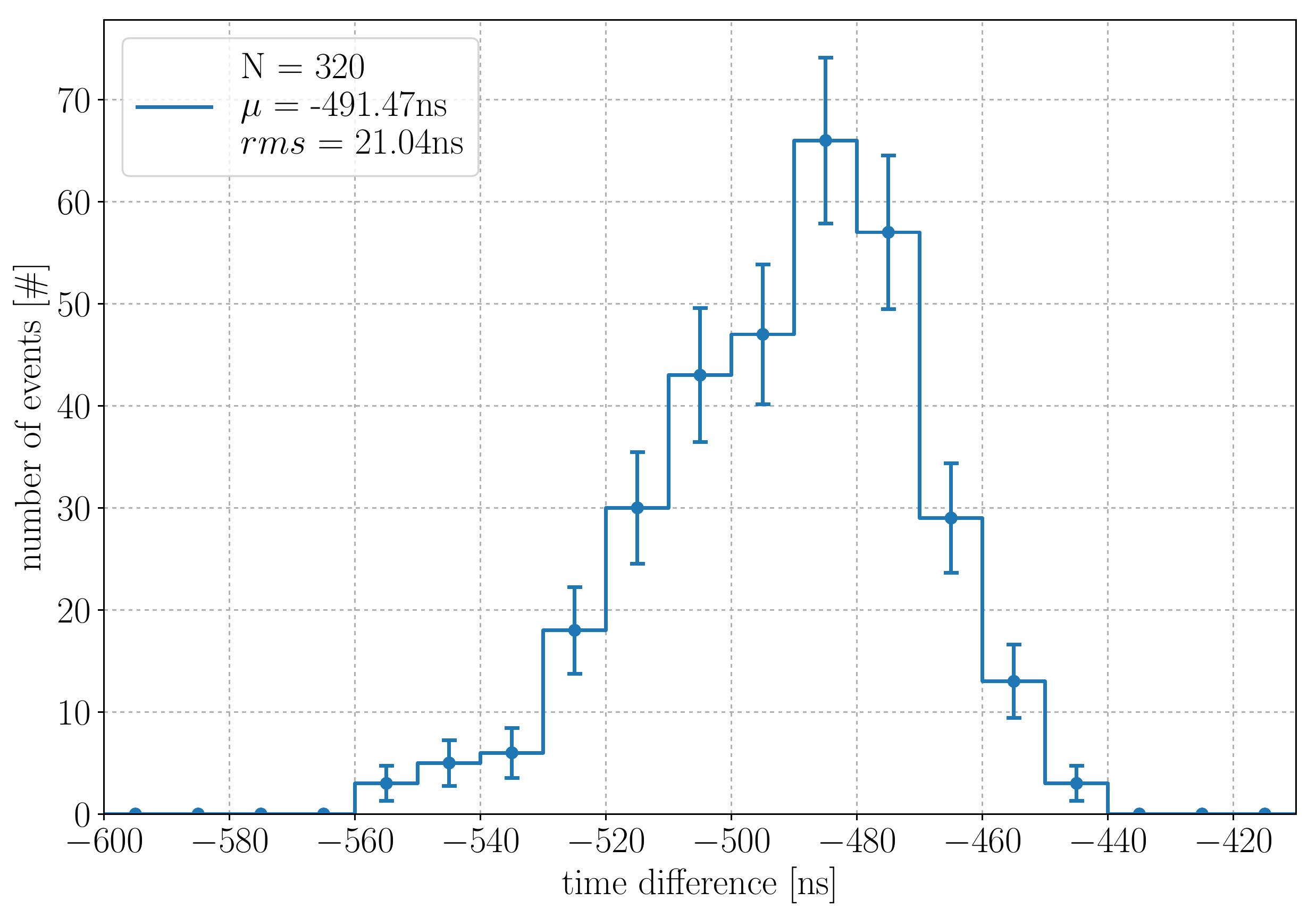}
		\subcaption{}
		\label{fig:deltat:icetop}
	\end{subfigure}
	\hfill
	\begin{subfigure}[t]{0.49\textwidth}
		\centering
		\includegraphics[width=\textwidth]{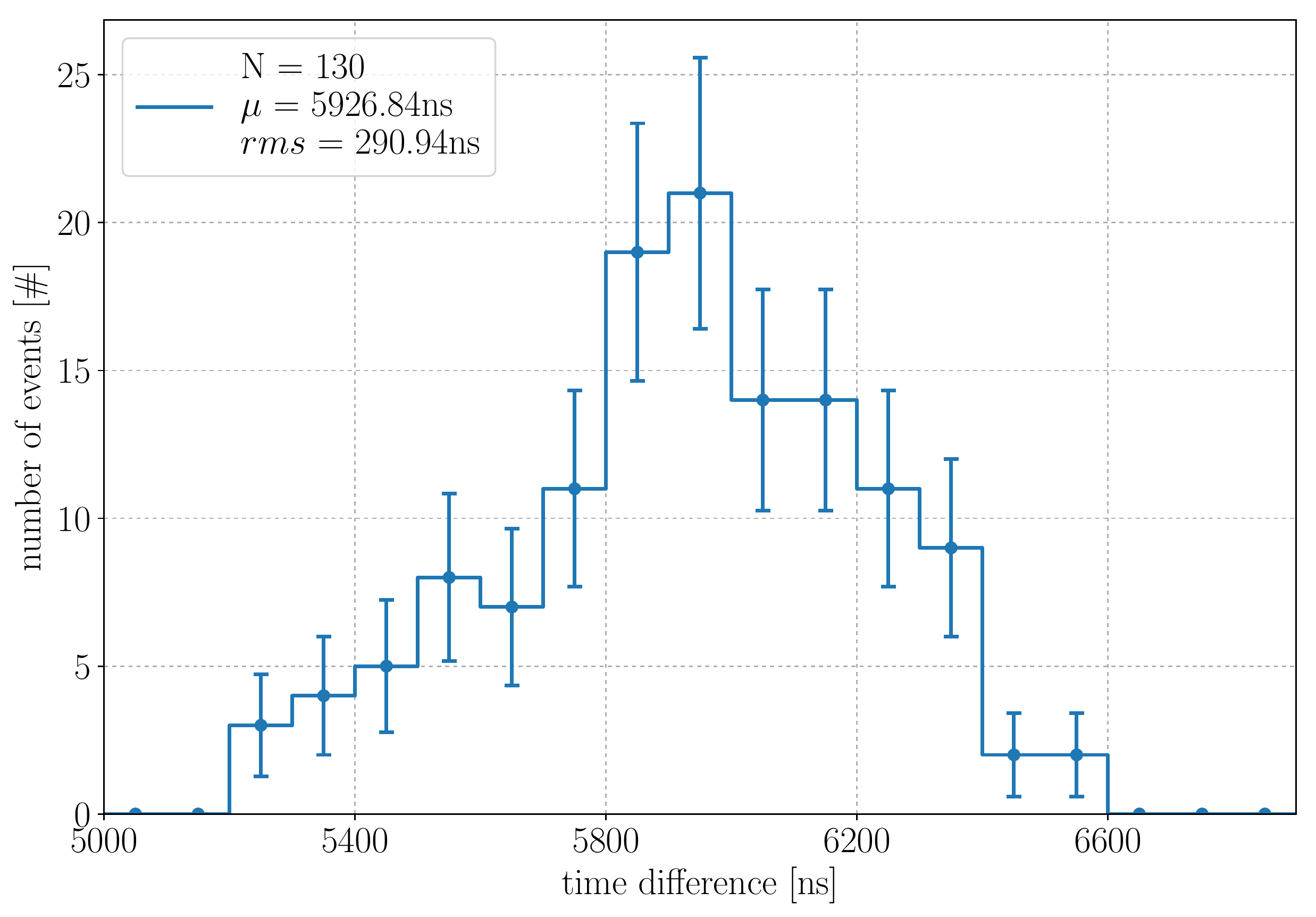}
		\subcaption{}
		\label{fig:deltat:icecube}
	\end{subfigure}
	\caption{Time difference distribution of the arrival time of the coincidence physics events recorded by: the IceAct and IceTop triggers (\subref{fig:deltat:icetop}) and the IceAct and IceCube in-ice triggers (\subref{fig:deltat:icecube}). Shown is the time difference $\Delta T = T_{\mathrm{IceTop/in-ice}} - T_{\mathrm{IceAct}}$.}
	\label{fig:deltat}
\end{figure}

Figure \ref{fig:deltat} shows the time difference between the IceTop or IceCube trigger and the arrival of the IceAct telescope trigger at the DOM main-board: $\Delta T = T_{\mathrm{IceTop/in-ice}} - T_{\mathrm{IceAct}}$. The arrival time of the IceAct trigger at the DOM main-board is delayed by a constant time offset due to the cable length connecting it with the IceAct telescope. This and an additional arbitrary software offset causes triggers of IceAct to arrive about \SI{490}{\nano \second} after IceTop trigger for coincident events as shown in Fig. \ref{fig:deltat:icetop}. For IceCube in-ice coincidences the events follow a clear event distribution given by the depth of the IceCube detector and the corresponding travel-time of the air shower particles. The time-correlation with respect to IceTop is better than \SI{21}{\nano\second}, with IceCube \SI{290}{\nano\second}, consistent with the expectation from geometrical smearing (especially for the IceCube in-ice array) and reconstruction uncertainties. The correlation demonstrates well the high purity of coincident detection.

\begin{figure}[htbp]
	\centering
	\includegraphics[width=.75\textwidth]{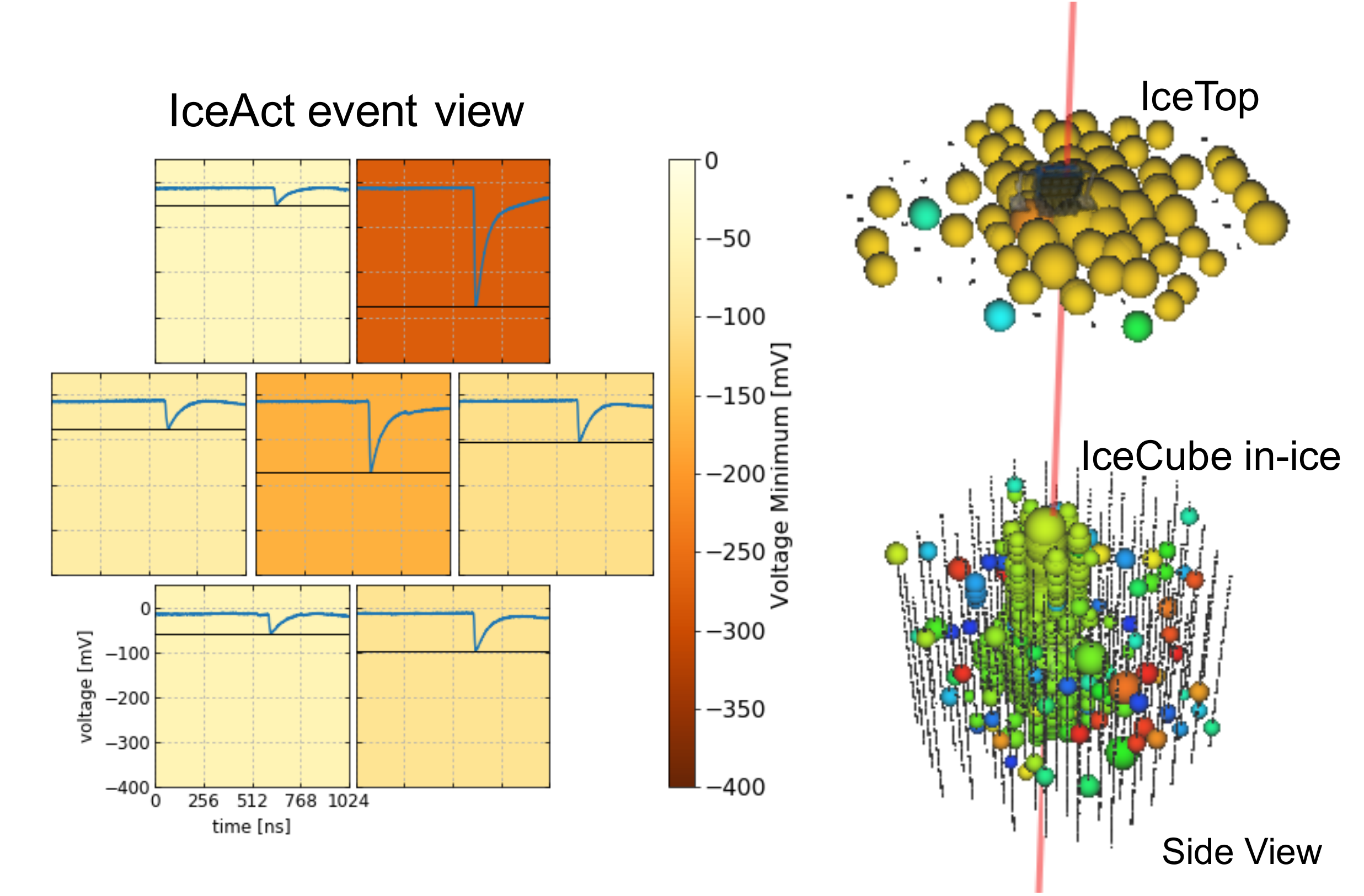}
	\caption{Example of a coincidence event. On the right side an IceCube event view of a vertical air shower in coincidence with the IceAct demonstrator is shown. The position and size of each bubble correspond to the position of the hit IceCube digital optical module and the amount of light it detected. The color indicates the time: from yellow for early hits to green for the later hits. Vertex, zenith and azimuth correspond to the in-ice muon track reconstruction indicated by the solid red line. The left plot shows the light pulses detected by the IceAct demonstrator in time-coincidence with this event. The black line indicates the detected pulse height of the signal.}
	\label{fig:examcoincidence}
\end{figure}

As an example, one of these coincident events is displayed as an event-view in Fig. \ref{fig:examcoincidence}.
A clear air-shower signal in the surface tanks of IceTop as well as the signal of a muon bundle in the IceCube detector can be seen. Signals from Cherenkov photons are visible in all seven pixels of the IceAct telescope. When propagated back to the surface, the reconstructed path of the muon bundle detected in IceCube coincides well with the position of IceAct on the rooftop of the IceCube Laboratory (ICL) at the center of the IceCube array.

\section{Performance based on the Analysis of Coincident Events}
\label{sec:performance}

In the following the performance of the IceAct demonstrator telescope will be discussed based on the data taken in coincidence with IceTop and IceCube.

\begin{figure}[tbp]
	\centering
	\includegraphics[width=.75\textwidth]{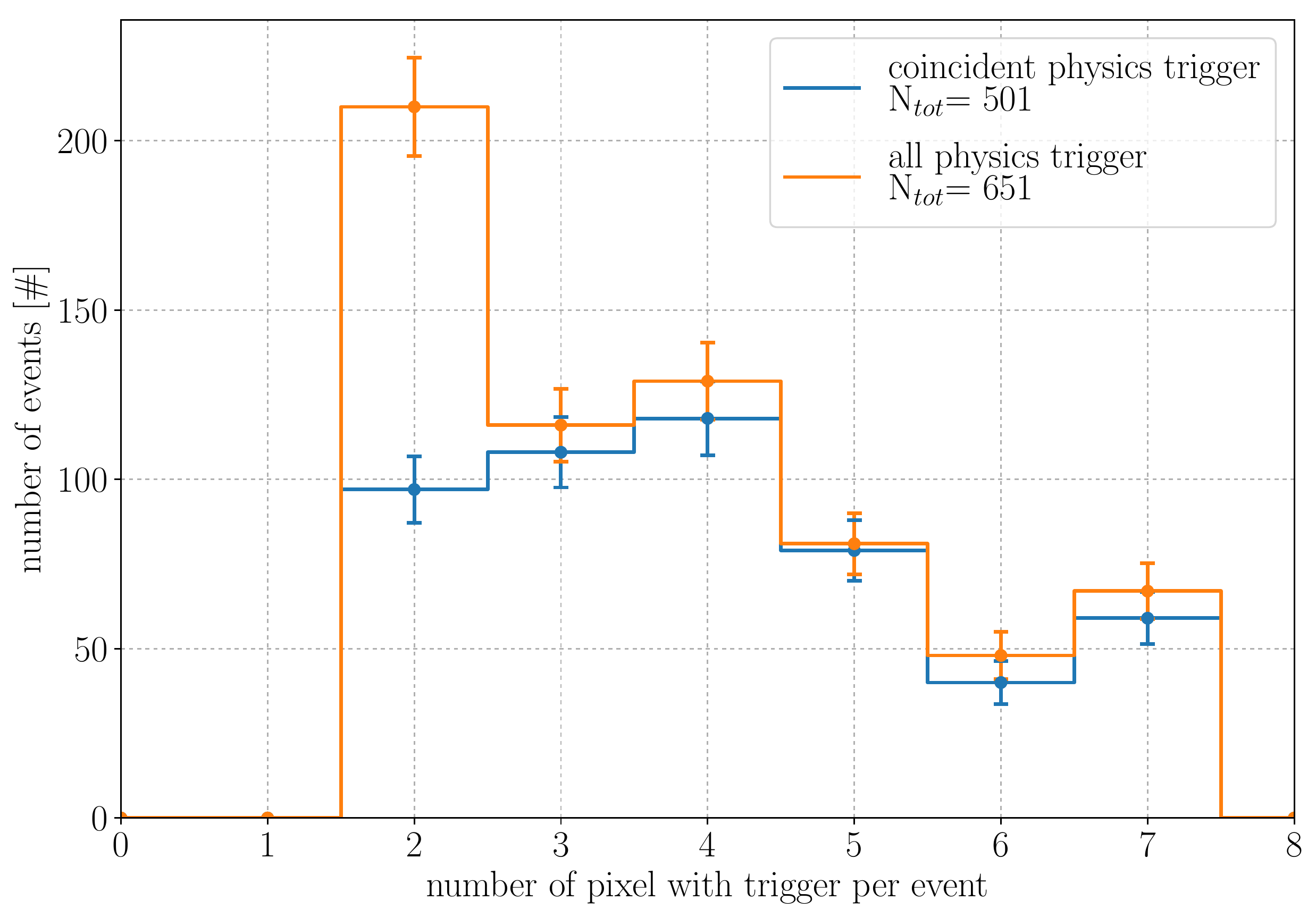}
	\caption{IceAct trigger multiplicity showing the number of pixels contributing to the trigger for all coincident physics events after all quality cuts.}
	\label{fig:pixelmultiplicity}
\end{figure}

\medskip
The pixel multiplicity of all selected IceAct, IceCube, and IceTop coincident events is shown in Fig. \ref{fig:pixelmultiplicity}. The software trigger requires two different pixels to have a signal over threshold. After this, single pixel events caused by the pixel that is connected twice to the DAQ system are excluded. It can be seen that the coincident events are not dominated by pixel numbers at the trigger threshold of $N_{pixel} = 2$. Compared to all physics triggers of the telescope, including non coincidences passing the same quality cuts, a clear excess towards larger pixel multiplicity is visible. Note that the bin $N_{pixel} = 7$ contains more entries as it acts as an overflow bin for high energy showers illuminating the full camera.

\begin{figure}[tbp]
	\centering
	\begin{subfigure}[t]{0.49\textwidth}
		\centering
		\includegraphics[width=\textwidth]{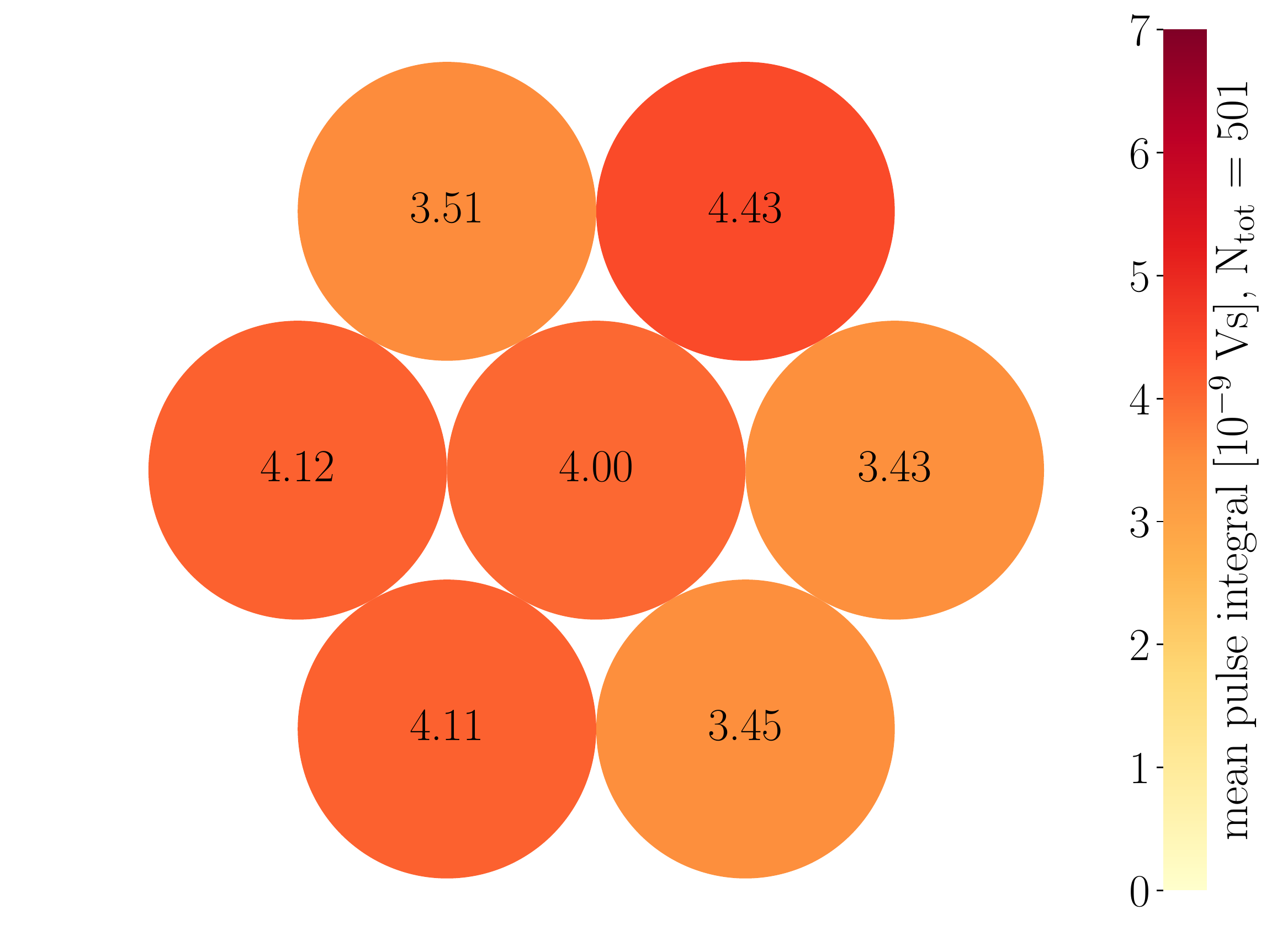}
		\subcaption[]{Mean pulse integral (charge equivalent) for each pixel of the IceAct telescope. The pulse heights are integrated over 64 ns after the time of the trigger.}
		\label{fig:pixelcharge}
	\end{subfigure}
	\hfill
	\begin{subfigure}[t]{0.49\textwidth}
		\centering
		\includegraphics[width=\textwidth]{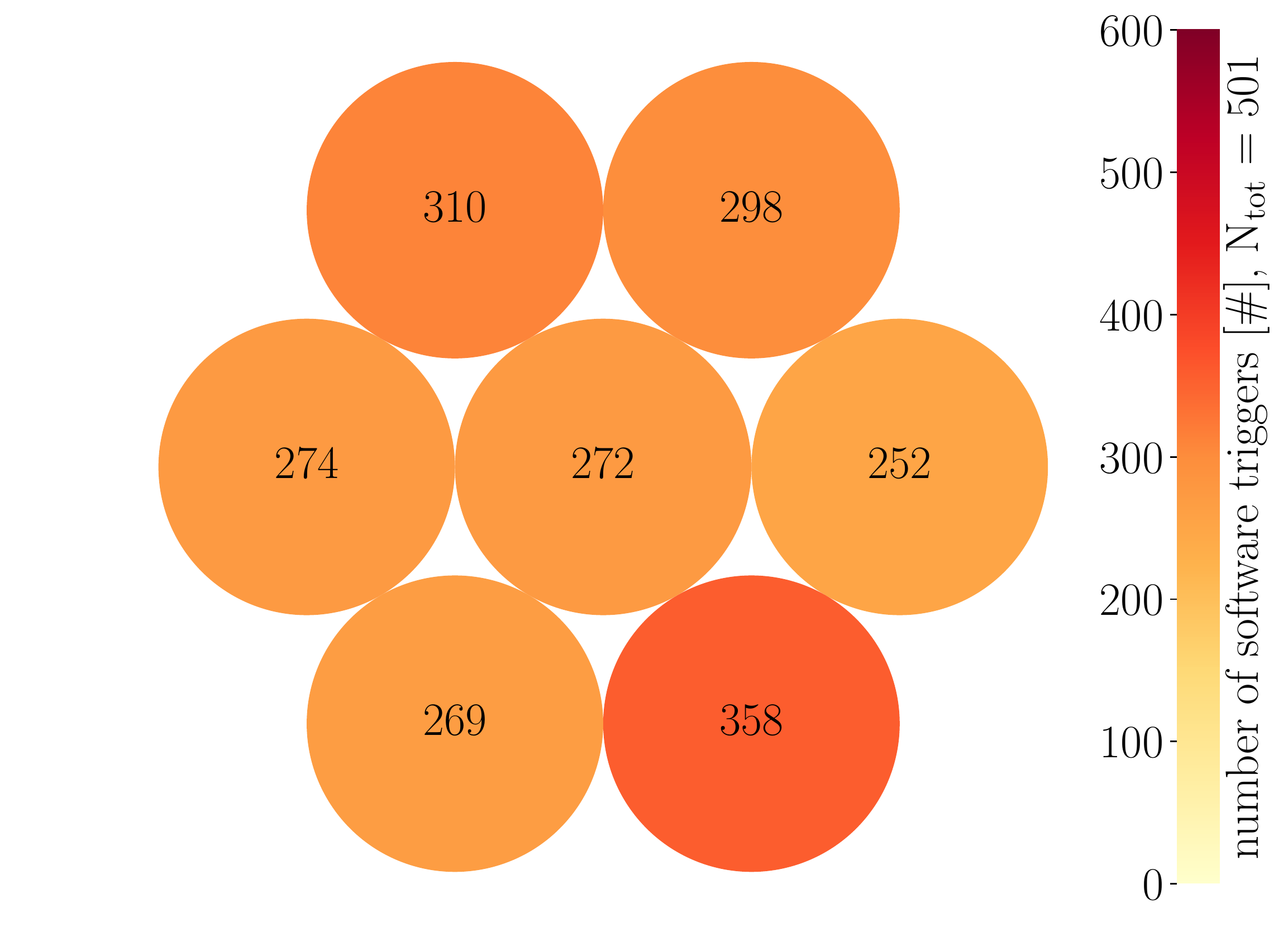}
		\subcaption[]{Number of events with a trigger signal in the corresponding pixel of IceAct for all physics triggered events.}
		\label{fig:pixeltriggger}
	\end{subfigure}
	\caption{Stability of the 7-pixel camera.}
\end{figure}

Figure~\ref{fig:pixelcharge} shows the mean pulse integral of each pixel within the coincident events in units of \SI{e-9}{Vs}. The integral is calculated over \SI{64}{\nano\second} starting at the trigger time found by the software trigger for each pixel. It shows a similar response of all pixels, even though the full demonstrator system was never completely calibrated for a pixel independent response to detected photons but only concerning a correct translation of ADC counts into voltage. Figure~\ref{fig:pixeltriggger} shows the number of trigger contributions from each pixel.

Both distributions, pixel-charge (Fig. \ref{fig:pixelcharge}) and pixel-trigger contribution (Fig. \ref{fig:pixeltriggger}), are compatible with a homogeneous distribution across the camera proving a reliable operation of all seven SiPMs during the run.

\begin{figure}[tbp]
	\centering
	\begin{subfigure}[t]{0.75\textwidth}
		\centering
		\includegraphics[width=\textwidth]{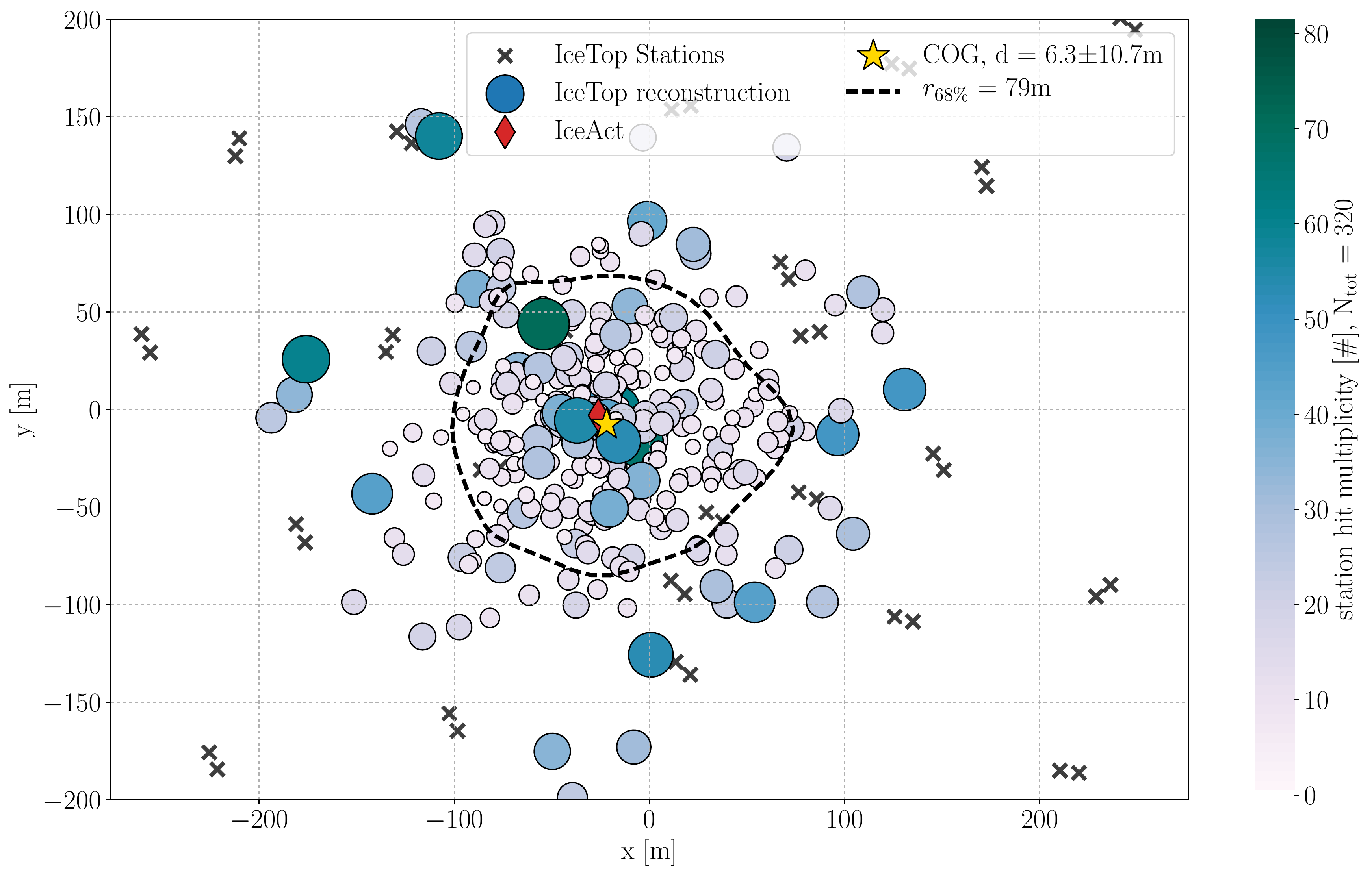}
		\subcaption[]{Size and color of the circles correspond to the IceTop station multiplicity.}
		\label{fig:icetop:core:NStation}
	\end{subfigure}
	\begin{subfigure}[t]{0.75\textwidth}
		\centering
		\includegraphics[width=\textwidth]{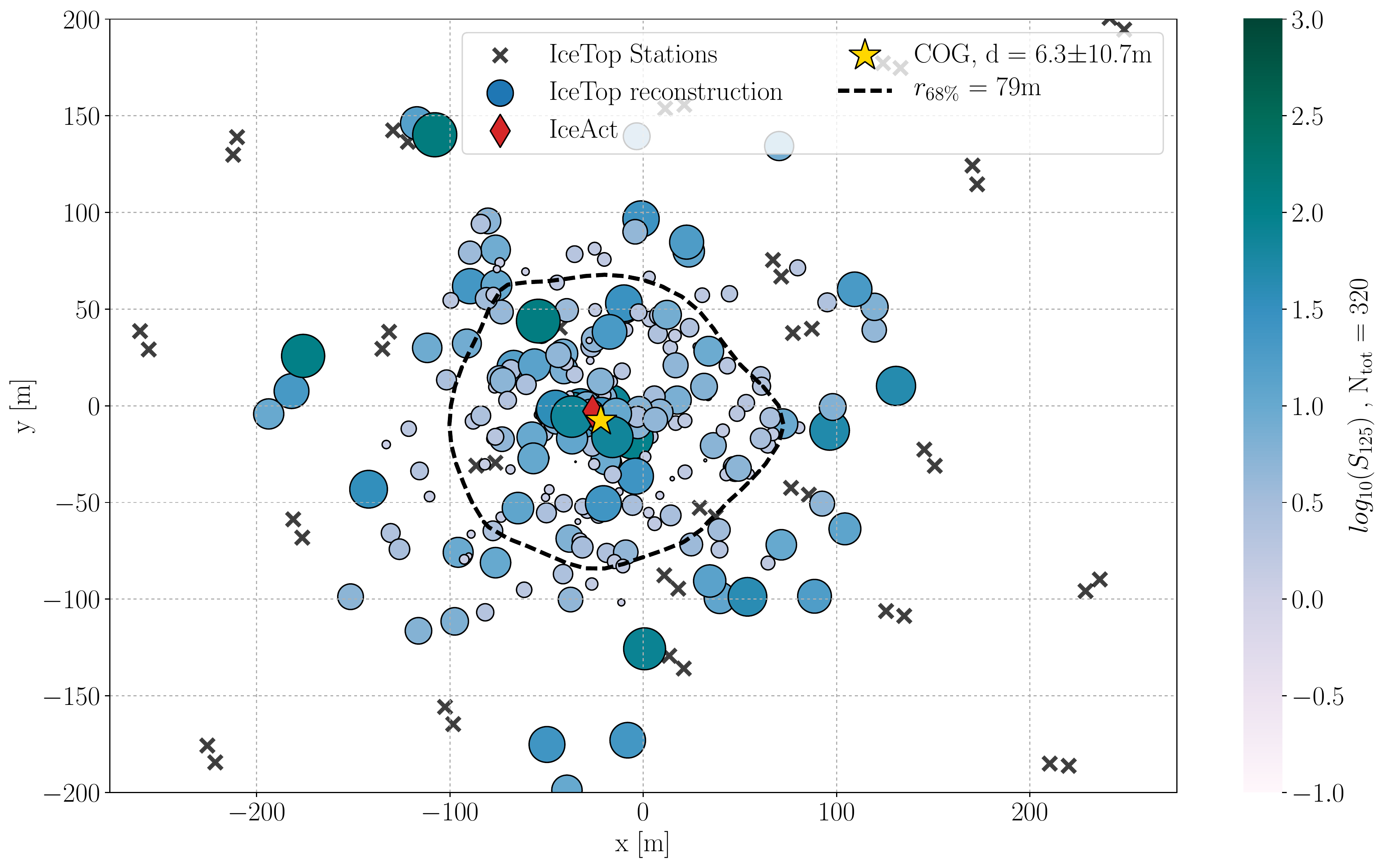}
		\subcaption[]{Size and color of the circles correspond to the IceTop $\log_{10}(S_{125})$ energy estimator.}
		\label{fig:icetop:core:S125}
	\end{subfigure}
	\caption{Air shower core positions as reconstructed by IceTop on the surface.}
	\label{fig:icetop:core}
\end{figure}

\begin{figure}[tbp]
	\centering
	\includegraphics[width=.75\textwidth]{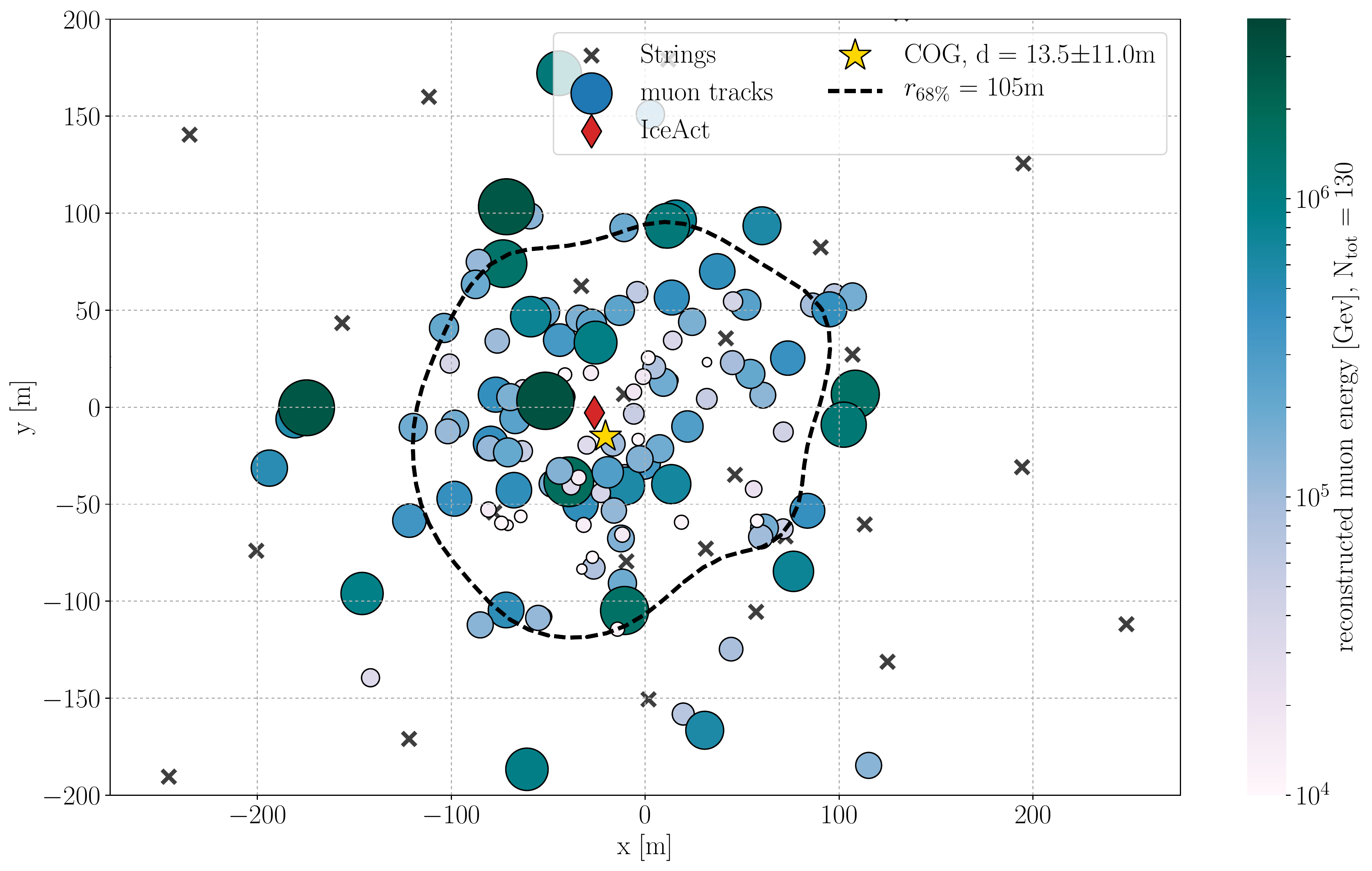}
	\caption{Position of the IceCube in-ice reconstructed muon back-tracked to the surface. Size and color of the circles correspond to the energy of the in-ice muon.}
    \label{fig:icecube:core}
\end{figure}

To prove the signals that were detected are indeed induced by cosmic rays and to illustrate the performance of the telescope, we analyzed the directional reconstructions of IceCube and IceTop for the coincident events. The x,y-distributions of the events that are triggered by IceAct and successfully reconstructed by IceCube and IceTop are clearly clustered around the IceAct demonstrator (see Figs.~\ref{fig:icetop:core} and \ref{fig:icecube:core}). The \SI{68}{\percent} contour shown in Figs.~\ref{fig:icetop:core} and \ref{fig:icecube:core} is constructed as the enclosing curve of the central \SI{68}{\percent} region of a two-dimensional Kernel Density Estimation of the corresponding impact points. The given radius corresponds to that of a circle, centered at IceAct, covering the same area.

The x,y distribution of the IceTop events show that \SI{68}{\percent} of the events cluster within a distances of up to about \SI{79}{m}. The mean of this distribution of events, called COG (Center Of Gravity), is \SI{6.3+-10.7}{m} away from the position of the telescope. Due to the long lever arm of the muon reconstruction deep in the ice, the x,y distribution of the muons reconstructed by IceCube is larger (Fig.~\ref{fig:icecube:core}). The IceCube events show that \SI{68}{\percent} of the events cluster within a radius of about \SI{105}{m} and the COG of the events is \SI{13.5+-11}{m} away from the position of the telescope. Both clearly indicate the coincident observation of Cherenkov light and muons originating from the same cosmic ray-induced air showers.
The color and size of each air-shower impact point, as reconstructed by
IceTop, represent the IceTop station hit multiplicity (Fig. \ref{fig:icetop:core:NStation}) as a reconstruction-independent and $S_{125}$ (Fig. \ref{fig:icetop:core:S125}) as a reconstruction-dependent energy estimator \cite{IceCube:2012nn} of that particular air-shower. Figure \ref{fig:icetop:core:NStation} clearly shows that small air showers are clustered closer to IceAct, while higher energy air-showers are observed in coincidence even if they are further away. The same conclusion holds for muons originating from air-showers, as shown in Fig.~\ref{fig:icecube:core}, where size and color corresponds to the energy reconstruction of the in-ice muon track. This energy-dependent impact point distribution is expected due to higher energy air-showers producing more Cherenkov light that spreads over a larger area.

\begin{figure}[tbp]
	\centering
	\includegraphics[width=.75\textwidth]{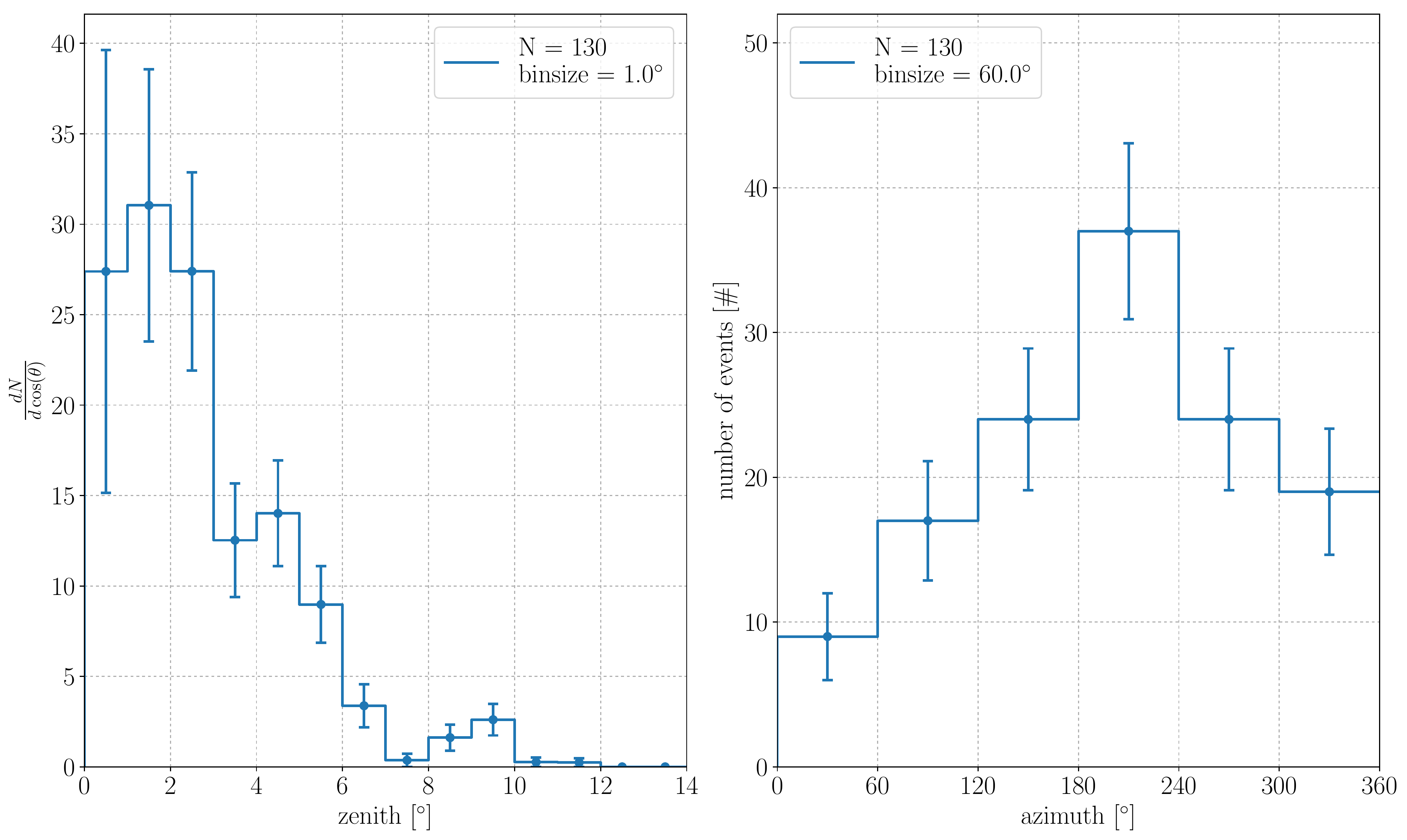}
	\caption{Zenith and azimuth distribution of IceAct-IceCube in-ice coincident events.}
	\label{fig:zenithazimuth:icecube}
\end{figure}

\begin{figure}[tbp]
	\centering
	\includegraphics[width=.75\textwidth]{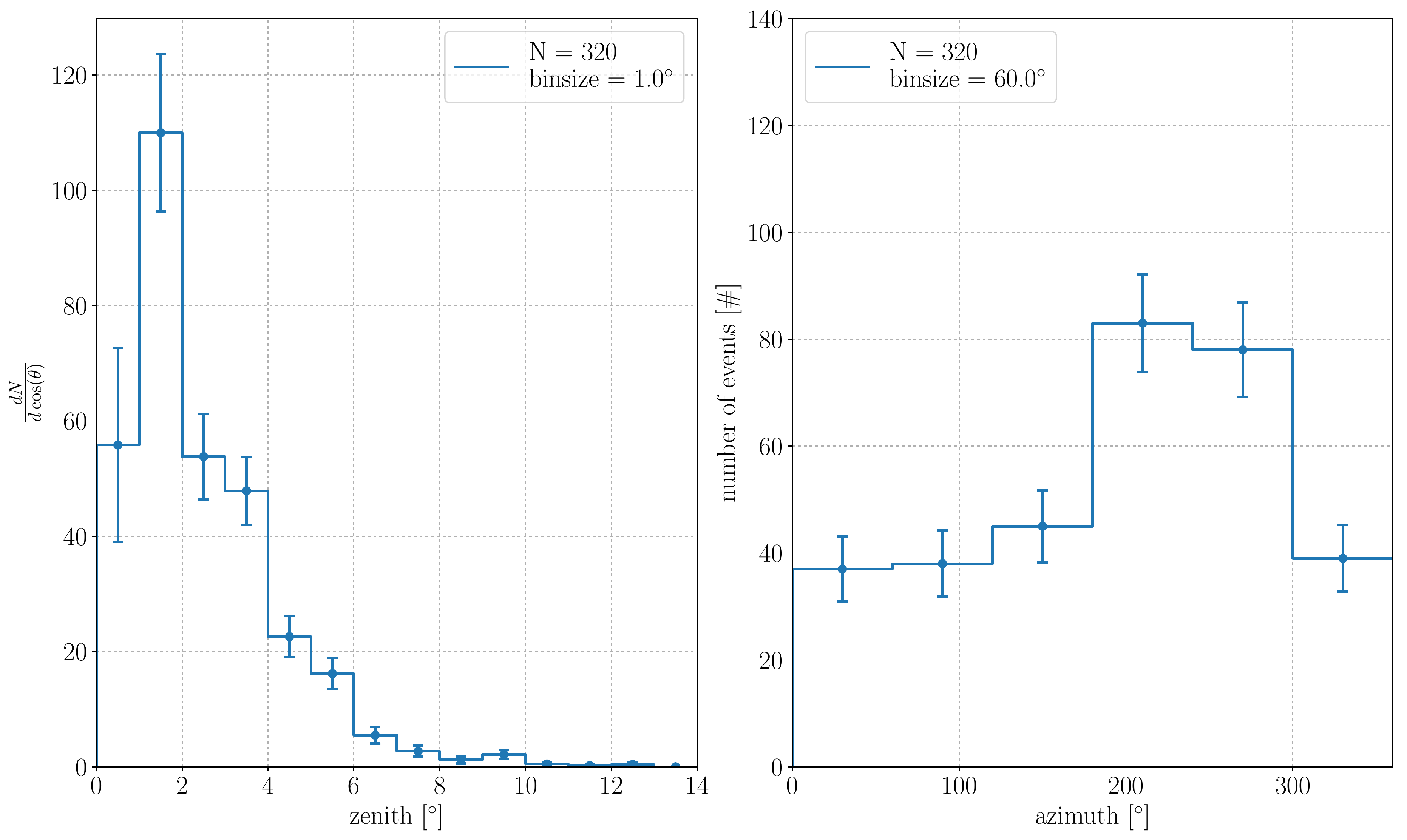}
	\caption{Zenith and azimuth distribution of IceAct-IceTop coincident events.}
	\label{fig:zenithazimuth:icetop}
\end{figure}

\begin{figure}[tbp]
	\centering
	\begin{subfigure}[t]{0.49\textwidth}
		\centering
		\includegraphics[width=\textwidth]{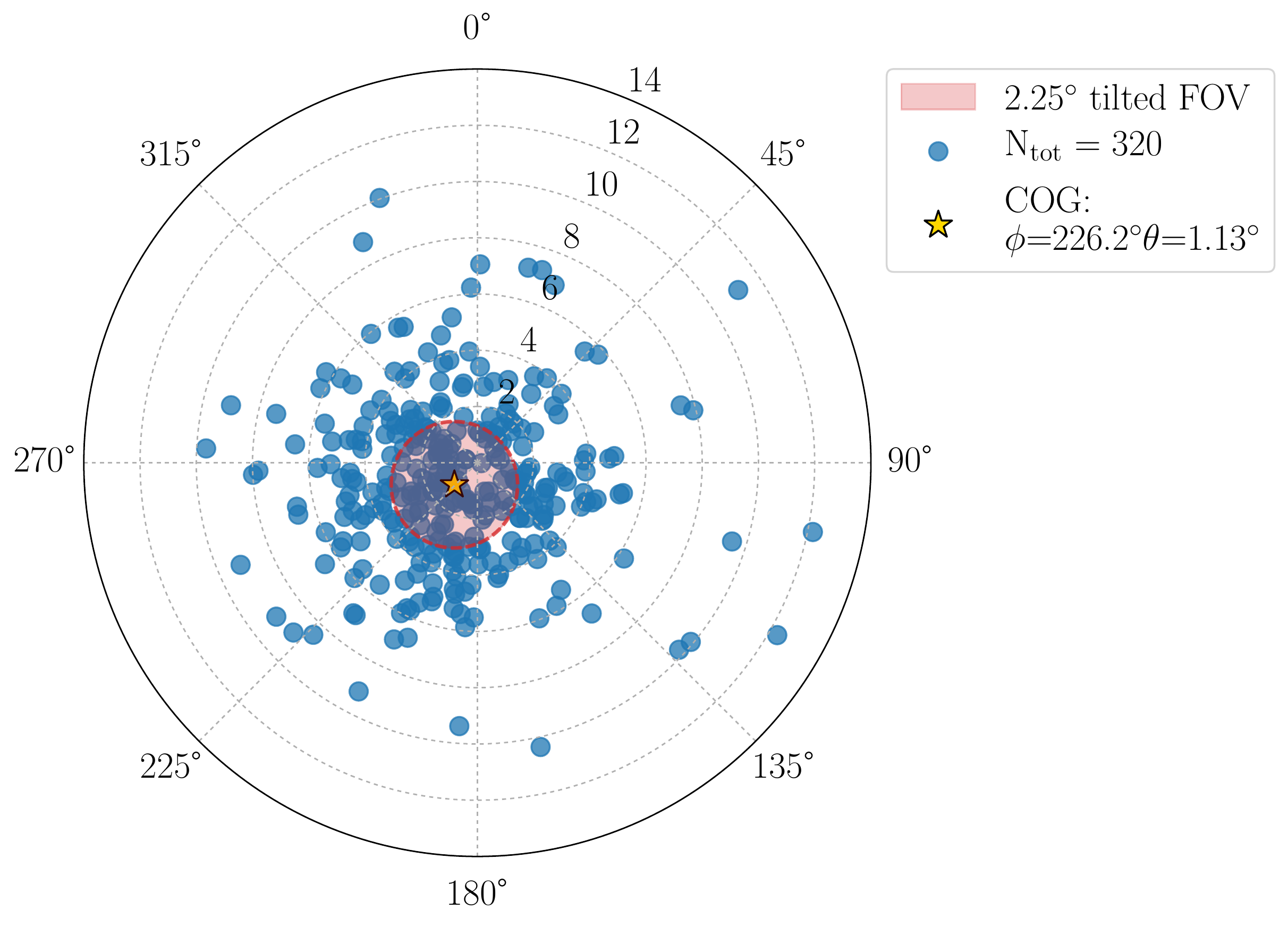}
		\subcaption[]{IceTop air showers}
		\label{fig:icetop:circdiagram}
	\end{subfigure}
	\hfill
	\begin{subfigure}[t]{0.49\textwidth}
		\centering
		\includegraphics[width=\textwidth]{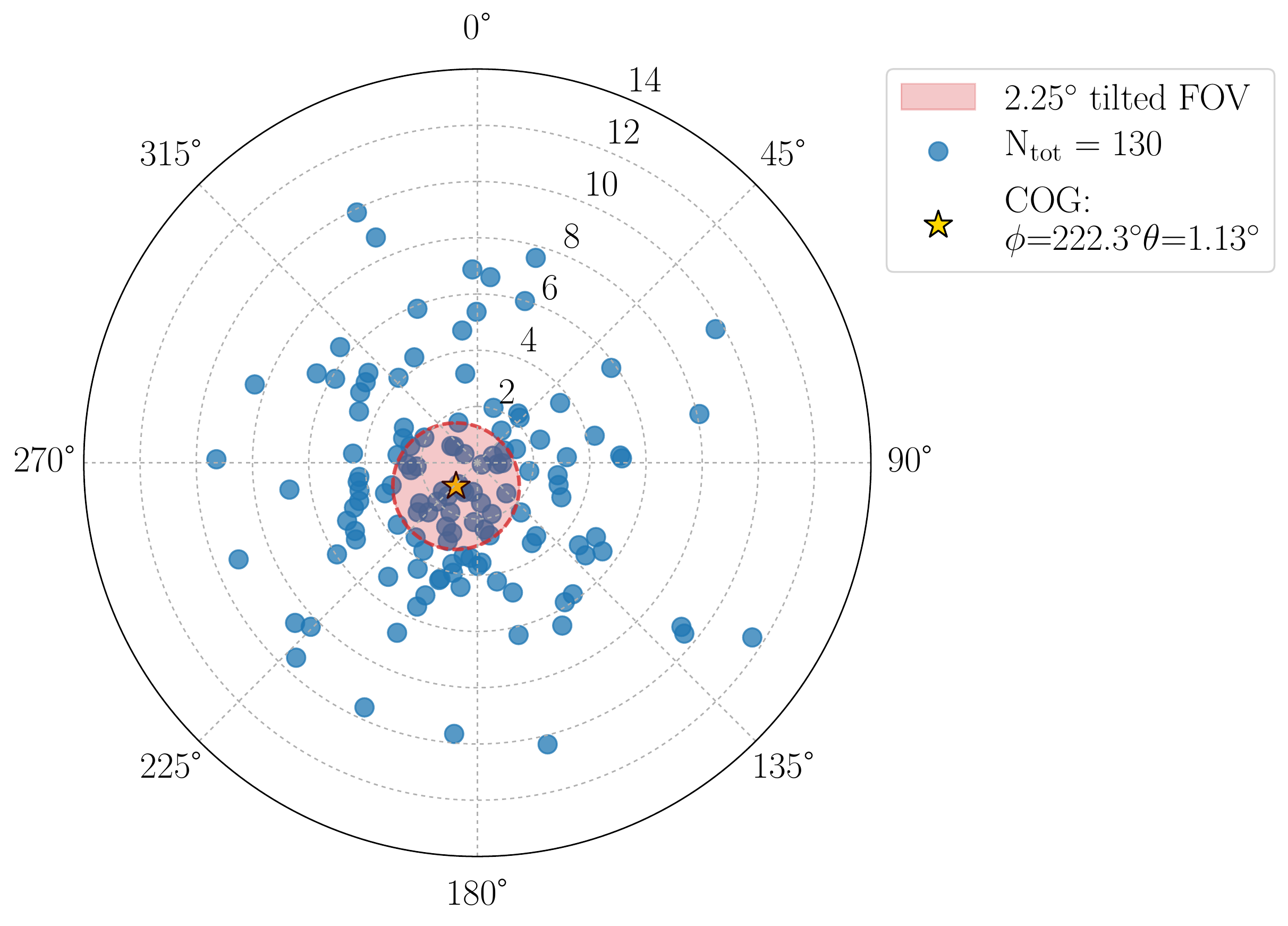}
		\subcaption[]{IceCube in-ice muons}
		\label{fig:icecube:circdiagram}
	\end{subfigure}
	\caption{Zenith and azimuth angle distribution of the the events measured with IceTop (\subref{fig:icetop:circdiagram}) and IceCube (\subref{fig:icecube:circdiagram}) that were in coincidence with IceAct. The red circle represents a \ang{2.25} field-of-view centered on the COG of the reconstructed event directions. It is offset from the vertical, $\theta = \ang{0}$, indicating the tilt of the telescope based on these reconstructions.}
	\label{fig:circdiagram}
\end{figure}

The directional reconstructions of the IceCube in-ice and IceTop events that are in coincidence with IceAct have a zenith angle distribution that peaks at around \ang{2} from the vertical due to the limited field-of-view of the 7-pixel telescope (see Figures \ref{fig:zenithazimuth:icecube} and \ref{fig:zenithazimuth:icetop}). We expect a field-of-view of about \ang{1.5} per pixel, resulting in a field-of-view with a diameter of about \ang{4.5} for the whole camera. Both azimuth distributions (right side of Figs. \ref{fig:zenithazimuth:icecube} and \ref{fig:zenithazimuth:icetop}) of the coincident events show a maximum around \ang{225}. This is due to a small shift of the telescope from the vertical direction, as shown in the circular diagrams of air-shower/muon reconstruction in Fig. \ref{fig:circdiagram}. In this polar $r-\phi$ representation, $r$ corresponds to the reconstructed zenith while $\phi$ is the azimuth. The distribution is clearly not centered but shifted into a direction of about \ang{225}. The tilt of the IceAct telescope is found to be in the order of \ang{1} as indicated by the COG in Fig. \ref{fig:circdiagram}. Both the events reconstructed based on the deep IceCube in-ice detector and the events reconstructed based on the IceTop detector, show the same shift (\ang{1.1}) in the same direction (~\ang{225}) and confirm the tilt of the telescope.

As a crosscheck, Monte Carlo events were randomly generated with a homogeneous distribution in azimuth and a zenith angle from  \ang{0} to \ang{2.25}. After that, the observation plane of the simulation was tilted in azimuth to agree with the observed distributions in order to confirm the tilt of the telescope. The telescope was found to be tilted by \ang{0.6} in zenith and \ang{230} in azimuth using fitting the Monte Carlo simulation to both, the data coincident with IceCube in-ice and independently the data in coincidence with IceTop. This calibration can generally be done very accurately with IceTop data. After the calibration of a possible tilt, future telescopes, equipped with full 61-pixel camera, can be used to improve the directional reconstruction of coincident events, especially in case of events with only limited information from IceTop.

\begin{figure}[tbp]
	\centering
	\begin{subfigure}[t]{0.49\textwidth}
		\centering
		\includegraphics[width=\textwidth]{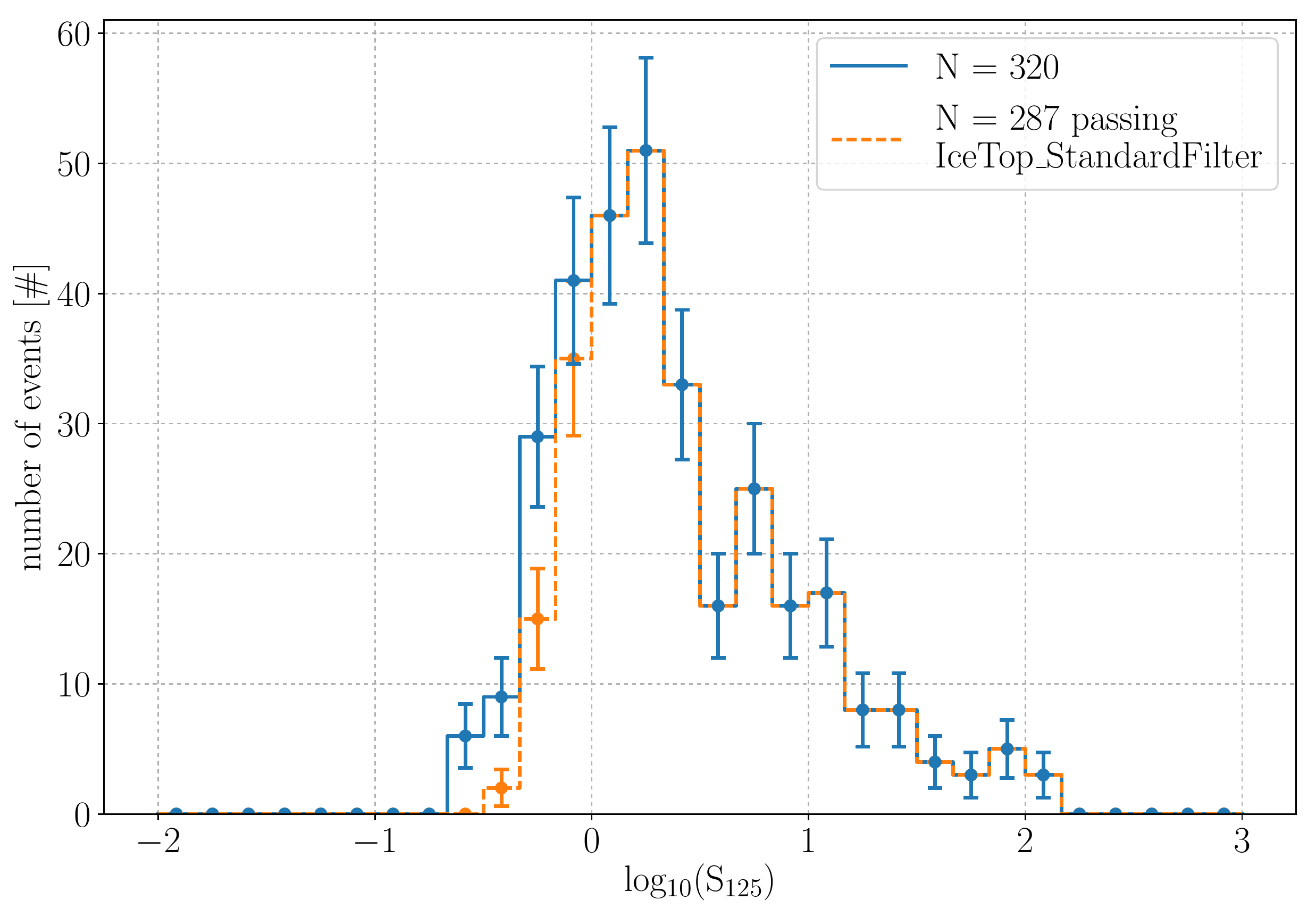}
		\subcaption[]{Distribution of IceTop $\log_{10}(S_{125})$ energy estimator. The blue curve is for all IceAct-IceTop coincidences while the dotted orange one shows only coincidences that would pass the additional quality cut of IceTop to ensure stable energy reconstruction. The additional events in the blue histogram that are not found in the orange histogram are those events where IceAct could improve IceTop's air shower reconstruction.}
		\label{fig:icetop:energy}
	\end{subfigure}
	\hfill
	\begin{subfigure}[t]{0.49\textwidth}
		\centering
		\includegraphics[width=\textwidth]{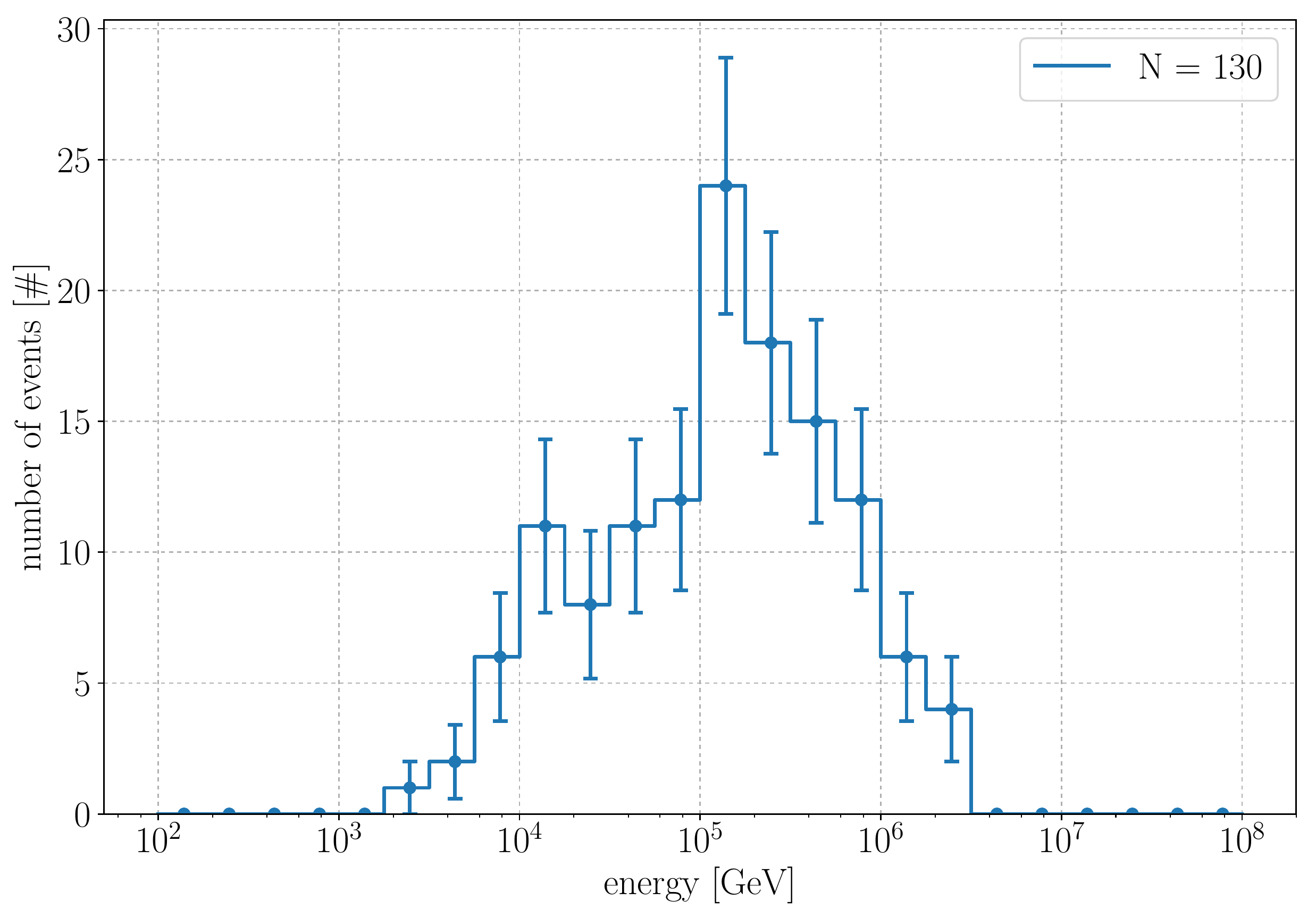}
		\subcaption[]{Reconstructed IceCube in-ice muon energy for IceAct coincident events. The energy range of observed events covers the region in which the telescope can function as a veto for atmospheric muons and neutrinos.}
		\label{fig:icecube:energy}
	\end{subfigure}
	\caption{Energy estimates of the events measured with IceTop (\subref{fig:icetop:energy}) and IceCube (\subref{fig:icecube:energy}) that were in coincidence with IceAct.}
\end{figure}

Based on standard reconstructions from IceCube \cite{IceCube:EnergyReco} and IceTop \cite{IceCube:2012nn}, a rough energy estimate of the in-ice muons and the primary energy based on the particles detected on the surface can be made (see Figures \ref{fig:icetop:energy} and \ref{fig:icecube:energy}). The shape of the energy distribution of IceTop events in coincidence with IceAct (Fig.~\ref{fig:icetop:energy}) shows a shape comparable to that of IceTop with additional IceTop standard quality cuts, indicating the energy-threshold of IceTop being dominant. For IceAct events in coincidence with IceCube, the peak of the event distribution of the IceAct events is clearly above that of IceCube in-ice only events indicating the IceAct energy threshold being dominant here (Fig.~\ref{fig:icecube:energy}). Future telescopes will have improved optics and may bring the energy threshold of the telescopes further down.

\section{Summary and Outlook}

This paper describes the technology of the first IceAct imaging air Cherenkov telescope demonstrator running during the astronomical night at the South Pole. It consists of a small camera based on SiPMs read out with DRS4 flash ADC boards. Despite changing weather conditions, the telescope was operated continuously over one dark South Pole season. The longest consecutive run of the telescope was analyzed and detected cosmic ray-induced events in coincidence with IceTop and the IceCube in-ice detector, proving the capability to run imaging air Cherenkov telescopes at the South Pole in coordination with the other experiments.

A slight tilt of the telescope was observed by analyzing the coincident data. This indicates that a calibration of the field-of-view of such telescopes will be possible, loosening the required precision of the positioning of the telescopes during deployment to the level of a degree.

A possible correlation of the IceAct events and the weather conditions is suggested by an analysis of data from the South Pole station of MPLNET. Snow accumulation on the telescope optics needs to be further investigated but was already now not preventing the telescope from continuously detecting cosmic rays. The overall duty cycle of an imaging air Cherenkov telescope array will need to be investigated further by also analyzing data from optical cameras with fish-eye optics.

As the next step, a first 61-pixel telescope was deployed at the South Pole in 2017 followed by another two telescopes deployed early 2019. These telescopes are intended to be used to calibrate the IceTop energy threshold, test snow removal and heating systems, and to further develop autonomous operation modes. A next step could be the deployment of a station of 7 IceAct telescopes.

The dominant electromagnetic part of the air showers measured by IceAct telescopes will allow to correlate the dependency of both the high energy muon component (IceCube) and the low energy muon component (IceTop) with the development of the electromagnetic part of the air showers in the atmosphere. This opens a window for composition analyses and the test of hadronic interaction models with a multi-component detector, especially in the energy range around the knee of the cosmic ray energy spectrum.

With a larger number of telescopes and a lower energy threshold, even the calibration of the light deposit of low energy muon bundles in the deep IceCube detector in correlation with the cosmic ray primary energy will be possible.

The technology of small imaging air-Cherenkov telescopes will be explored as a potential detector component of the multi component detector IceCube and IceCube-Gen2. Already a small number of telescopes will be able to measure the electromagnetic part of air-showers in coincidence with IceCube and IceTop and will thus be able to calibrate the energy scale of these systems for cosmic ray detection, reducing for example systematic uncertainties due to snow accumulation.

The scientific goal of a large array of these small imaging air-Cerenkov telescopes is the precise observation of the development of the electromagnetic part air shower for cosmic ray and gamma ray detection in combination with IceTop, IceCube, and other future components like an array of scintillator panels. In addition an efficient detection of air showers can be used to veto cosmic rays in order to improve the detection and identification of astrophysical neutrino signals in the southern sky with the IceCube detector deep in the ice.

\acknowledgments

The IceCube collaboration acknowledges the significant contributions to this manuscript from Jan Auffenberg and Erik Ganster. The authors gratefully acknowledge the support from the following agencies and institutions:
USA {\textendash} U.S. National Science Foundation-Office of Polar Programs,
U.S. National Science Foundation-Physics Division,
Wisconsin Alumni Research Foundation,
Center for High Throughput Computing (CHTC) at the University of Wisconsin-Madison,
Open Science Grid (OSG),
Extreme Science and Engineering Discovery Environment (XSEDE),
U.S. Department of Energy-National Energy Research Scientific Computing Center,
Particle astrophysics research computing center at the University of Maryland,
Institute for Cyber-Enabled Research at Michigan State University,
and Astroparticle physics computational facility at Marquette University;
Belgium {\textendash} Funds for Scientific Research (FRS-FNRS and FWO),
FWO Odysseus and Big Science programmes,
and Belgian Federal Science Policy Office (Belspo);
Germany {\textendash} Bundesministerium f{\"u}r Bildung und Forschung (BMBF),
Deutsche Forschungsgemeinschaft (DFG),
Helmholtz Alliance for Astroparticle Physics (HAP),
Initiative and Networking Fund of the Helmholtz Association,
Deutsches Elektronen Synchrotron (DESY),
and High Performance Computing cluster of the RWTH Aachen;
Sweden {\textendash} Swedish Research Council,
Swedish Polar Research Secretariat,
Swedish National Infrastructure for Computing (SNIC),
and Knut and Alice Wallenberg Foundation;
Australia {\textendash} Australian Research Council;
Canada {\textendash} Natural Sciences and Engineering Research Council of Canada,
Calcul Qu{\'e}bec, Compute Ontario, Canada Foundation for Innovation, WestGrid, and Compute Canada;
Denmark {\textendash} Villum Fonden, Danish National Research Foundation (DNRF), Carlsberg Foundation;
New Zealand {\textendash} Marsden Fund;
Japan {\textendash} Japan Society for Promotion of Science (JSPS)
and Institute for Global Prominent Research (IGPR) of Chiba University;
Korea {\textendash} National Research Foundation of Korea (NRF);
Switzerland {\textendash} Swiss National Science Foundation (SNSF);
United Kingdom {\textendash} Department of Physics, University of Oxford.

The MPLNET project is funded by the NASA Radiation Sciences Program and Earth Observing System. We thank NOAA staff for operating and maintaining the MPLNET lidar at the South Pole. Logistical support for this project in Antarctica was provided by the U.S. National Science Foundation through the U.S. Antarctic Program.

\bibliographystyle{JHEP}
\bibliography{references}

\end{document}